\documentclass[preprint2]{aastex631}
\usepackage{newtxtext,newtxmath}

\usepackage{bm,enumitem}

\begin{document}

\title{MACER3D --- an upgrade of MACER2D with enhanced subgrid models and gas physics --- and its application to simulating AGN feedback in a massive elliptical galaxy}

\correspondingauthor{Suoqing Ji, Feng Yuan}
\email{sqji@fudan.edu.cn, fyuan@fudan.edu.cn}

\author[0009-0008-0460-8685]{Haoen Zhang}
\altaffiliation{These authors contributed equally to this work.}
\affiliation{Astrophysics Division, Shanghai Astronomical Observatory, Chinese Academy of Sciences, 80 Nandan Road, Shanghai 200030, P.R.China}
\affiliation{University of Chinese Academy of Sciences, No. 19A Yuquan Road, Beijing 100049, P.R.China}

\author[0009-0004-3881-674X]{Haojie Xia}
\altaffiliation{These authors contributed equally to this work.}
\affiliation{Astrophysics Division, Shanghai Astronomical Observatory, Chinese Academy of Sciences, 80 Nandan Road, Shanghai 200030, P.R.China}
\affiliation{University of Chinese Academy of Sciences, No. 19A Yuquan Road, Beijing 100049, P.R.China}

\author[0000-0001-9658-0588]{Suoqing Ji}
\affiliation{Center for Astronomy and Astrophysics and Department of Physics, Fudan University, Shanghai 200438, P.R.China}
\affiliation{Key Laboratory of Nuclear Physics and Ion-Beam Application (MOE), Fudan University, Shanghai 200433, P.R.China}

\author[0000-0003-3564-6437]{Feng Yuan}
\affiliation{Center for Astronomy and Astrophysics and Department of Physics, Fudan University, Shanghai 200438, P.R.China}

\author[0009-0009-7849-2643]{Minhang Guo}
\affiliation{Astrophysics Division, Shanghai Astronomical Observatory, Chinese Academy of Sciences, 80 Nandan Road, Shanghai 200030, P.R.China}
\affiliation{ShanghaiTech University, 393 Middle Huaxia Road, Shanghai 201210, P.R.China}
\affiliation{University of Chinese Academy of Sciences, No. 19A Yuquan Road, Beijing 100049, P.R.China}

\author[0009-0005-6382-2532]{Rui Zhang}
\affiliation{Astrophysics Division, Shanghai Astronomical Observatory, Chinese Academy of Sciences, 80 Nandan Road, Shanghai 200030, P.R.China}
\affiliation{University of Chinese Academy of Sciences, No. 19A Yuquan Road, Beijing 100049, P.R.China}

\author[0000-0003-0900-4481]{Bocheng Zhu}
\affiliation{National Astronomical Observatories, Chinese Academy of Sciences, 20A Datun Road, Beijing 100101, P.R.China}

\author[0000-0003-1717-6697]{Yihuan Di}
\affiliation{Department of Astronomy, Shanghai Jiao Tong University, 800 Dongchuan Road, Shanghai 200240, P.R.China}

\author[0000-0002-4882-8953]{Aoyun He}
\affiliation{Astrophysics Division, Shanghai Astronomical Observatory, Chinese Academy of Sciences, 80 Nandan Road, Shanghai 200030, P.R.China}
\affiliation{University of Chinese Academy of Sciences, No. 19A Yuquan Road, Beijing 100049, P.R.China}

\author[0009-0003-1442-105X]{Tingfang Su}
\affiliation{Astrophysics Division, Shanghai Astronomical Observatory, Chinese Academy of Sciences, 80 Nandan Road, Shanghai 200030, P.R.China}
\affiliation{University of Chinese Academy of Sciences, No. 19A Yuquan Road, Beijing 100049, P.R.China}

\author[0009-0006-4662-3053]{Yuxuan Zou}
\affiliation{Astrophysics Division, Shanghai Astronomical Observatory, Chinese Academy of Sciences, 80 Nandan Road, Shanghai 200030, P.R.China}
\affiliation{University of Chinese Academy of Sciences, No. 19A Yuquan Road, Beijing 100049, P.R.China}

\begin{abstract}
    We present MACER3D (Multiscale AGN-regulated Cosmic Ecosystem Resolver in 3D), a new suite of three-dimensional hydrodynamic simulations that study active galactic nuclei (AGN) feedback on galactic scales over Gyr duration, with major enhancement in subgrid models and gas physics over its predecessor -- MACER (Massive AGN Controlled Ellipticals Resolved) which is in two dimensions (hereafter MACER2D). MACER3D resolves gas dynamics from within the Bondi radius ($\sim 25\,\mathrm{pc}$) to halo scales. Combined with black hole accretion theory, it enables an accurate calculation of AGN outputs and subsequently their large-scale feedback effects. We present results from simulating an isolated elliptical galaxy with different feedback configurations. In the fiducial model with both AGN and supernova (SN) feedback, the temporal evolution of AGN luminosity and star formation rate are strongly correlated, suggesting shared dependence on the availability of gas supply for SMBH accretion and star formation. AGN duty cycles of several percent with a single-cycle timescale of $\sim 10^2\,\mathrm{Myr}$ agree with observations, while models with only AGN or SN feedback fail to reproduce observed cycles. While all models maintain a quiescent galaxy state, fiducial AGN+SN feedback model results in higher star formation than no-SN feedback, suggesting SN feedback, when acting synergistically with AGN feedback, may positively impact star formation. Combined AGN and SN feedback enhances halo-scale metal enrichment compared to single-feedback models. The simulated X-ray properties match observations and predict transient cavities produced by cold-mode AGN winds from past burst events. The differences between the results obtained by MACER2D and MACER3D are also discussed.
\end{abstract}

\keywords{galaxies: formation --- galaxies: evolution --- methods: numerical --- hydrodynamics}

\section{Introduction} \label{sec:intro}

The interaction between supermassive black holes (SMBHs) and their host galaxies represents a fundamental aspect of galaxy evolution \citep{somerville15, naab17, crain23}. Active galactic nuclei (AGN) feedback, which encompasses the energy and momentum injected into the interstellar medium (ISM) and circumgalactic medium (CGM) by the central SMBH, plays an indispensable role in regulating the growth of both galaxies and SMBHs, as well as shaping the properties of galaxies and their environments \citep{tumlinson2017circumgalactic,faucher2023key}. AGN feedback is considered responsible for several observed scaling relations between SMBHs and their host galaxies, including the $M_\mathrm{BH}-\sigma$ relation, the $M_\mathrm{BH}-M_\mathrm{bulge}$ relation, and the $M_\mathrm{BH}-L_\mathrm{bulge}$ relation \citep{magorrian98, tremaine02, gebhardt00, ferrarese00, haring04, guktekin2009, kormendy&ho13, zhuang23}. Moreover, AGN feedback contributes significantly to the quenching of star formation in massive galaxies, the suppression of cooling flows in galaxy clusters, and the heating of the CGM \citep{croton06, li15, su21, zhu23a}.

Significant research efforts over the past few decades have focused on understanding the physical processes of AGN feedback and its impact on galaxy evolution \citep[etc.]{dimatteo05, springel05, hopkins06, croton06, sijacki07, gaspari12, li15, zinger20, su21}. However, substantial challenges persist in the theoretical modeling of these feedback processes and in reconciling theoretical predictions with observations. A primary challenge stems from the inherently multi-scale nature of AGN feedback, which spans a vast range of spatial and temporal scales - from the accretion disk around the SMBH to the large-scale environment of the galaxy and its CGM. The AGN duty cycle, representing the episodic nature of black hole accretion and feedback, involved with timescales ranging from brief bursts of $0.1\,\mathrm{Myr}$ to extended active phases approaching $\sim\mathrm{Gyr}$. This variability reflects the complex interplay between gas availability and accretion processes on parsec scales, coupled with feedback mechanisms and CGM dynamics on kiloparsec scales and beyond, which collectively regulate black hole growth and star formation in host galaxies.

Beyond the scale-related challenges, AGN feedback manifests in diverse forms determined by the accretion rate of the central SMBH: the cold (or quasar/radiative) mode and the hot (or radio/kinetic) mode \citep{fabian12}. The cold mode, occurring when the SMBH accretes near and above the Eddington rate, is characterized by intense radiation and powerful winds \citep{murray95, bottorff97}. This mode, typically associated with luminous quasars, drives large-scale outflows that heat and expel gas from the galaxy, thereby suppressing star formation \citep{springel05, hopkins06}. Conversely, the hot mode, associated with lower accretion rates, generates relativistic jets and weaker winds \citep{yuan14}. This mode, common in low-luminosity AGNs, inflates bubbles and cavities in the hot gas of galaxy clusters, preventing gas cooling and subsequent star formation \citep{li15, su21}. The feedback energy, originating from gas accretion onto the SMBH, manifests through various mechanisms including kinetic energy of jets and winds \citep{yuan15}, turbulent heating, shock thermalization \citep{bambic}, radiation pressure \citep{costa18a}, and non-thermal processes such as magnetic fields \citep{cen24} and cosmic rays (CRs) \citep{su21}. These feedback processes exhibit highly nonlinear behavior and couple with other feedback mechanisms, influencing jet and wind launching, ISM/CGM interactions, turbulence generation, and multiphase gas mixing.

The complexity arising from this diversity in scales and physics poses significant challenges for comprehensive modeling of AGN feedback, particularly in cosmological and zoom simulations of galaxy formation and evolution. Sub-grid models typically address unresolved AGN feedback processes, including gas accretion onto the SMBH, jet and wind launching, and feedback energy/momentum interactions with the ISM and CGM. However, implementation approaches vary substantially across different simulations \citep{vogelsberger14, dubois14, crain15, weinberger17, dave19, wellons23, hopkins23}. For instance, IllustrisTNG incorporates a kinetic mode at low accretion rates and a thermal mode at high accretion rates \citep{weinberger17}, while EAGLE employs a thermal feedback model with fixed efficiency \citep{crain15}. SIMBA combines both kinetic and thermal feedback, emphasizing kinetic feedback at low accretion rates \citep{dave19}. FIRE-3 implements a comprehensive model including radiation, mechanical winds, and cosmic rays \citep{wellons23,hopkins23}. Despite these variations in implementation, consensus exists regarding AGN feedback's role in suppressing star formation and quenching massive galaxies through gas heating or outflow-driven gas removal.

Despite extensive observational evidence of AGN activity across various scales, the detailed physical processes of AGN feedback and its evolutionary impact remain incompletely understood. Notably, limited observational evidence supports instantaneous, negative feedback effects from AGNs \citep{shin19}. Some observations suggest positive AGN feedback, indicated by correlations between AGN luminosity and host galaxy star formation rates -- an apparent contradiction to theoretical expectations of AGN-induced star formation quenching \citep{Cresci15}. On the other hand, some simulations indicate that AGN feedback may locally enhance star formation through gas compression \citep{gaibler12, mf23}. Overall, the integrated impacts of AGN feedback on global star formation remains debated.

Addressing these challenges requires implementing a comprehensive, multi-dimensional, and multi-physics model capturing essential AGN feedback processes, particularly precise determination of central SMBH accretion rates that govern the AGN power and concrete AGN outputs. Motivated by this need, \citet{yuan18} developed the Massive AGN Controlled Ellipticals Resolved (MACER, hereafter MACER2D) project - a two-dimensional axisymmetric hydrodynamic framework studying the evolution of elliptical galaxies with the effects of AGN feedback included. MACER2D has several key features. It focuses on galactic rather than cosmological scales thus it has achieved very high spatial resolution. In fact, the Bondi radius, which is the outer boundary of the accretion flow of the central AGN, is well resolved. In this case, the mass flux within the Bondi radius can be accurately calculated. Combined with the black hole accretion theory, the mass accretion rate at the black hole horizon and the accordingly computed AGN outputs can be obtained, including the AGN power and properties of radiation, jet, and wind. These quantities are crucial for the study of AGN feedback. Moreover, the interaction between these outputs and ISM is calculated rather than parameterized as in almost all cosmological simulations. Admittedly, we note that the model is still idealized and misses the proper cosmological context, which is discussed in more detail in the caveat of the conclusions. In addition, although the mass accretion rate can be obtained more reliably than in cosmological simulations, the effects of the feedback are still parameterized in a subgrid fashion based on studies of accretion disk physics. MACER2D has studied the role of AGN feedback in the evolution of massive slow rotators \citep{yuan18}, massive fast rotators \citep{yoon18}, compact galaxies \citep{di23} and disk galaxies (Zou et al. in prep), ISM properties of massive galaxies \citep{li18}, the impact and fate of cosmological inflow in elliptical galaxies \citep{zhu23a}, the impacts of different modes of AGN feedback \citep{yoon19,zhu23b}, and the effects of parameter variation of AGN feedback \citep{yao21}.

Recent advances in computational capabilities and improved understanding of feedback microphysics have enabled increasingly feasible and necessary three-dimensional AGN feedback simulations \citep{hopkins24b,hokpins24a, guo23}. These simulations more accurately capture turbulence effects, instabilities, and non-axisymmetric structures crucial for understanding multiphase gas mixing in the ISM and CGM. Consequently, we have developed the Multiscale AGN-regulated Cosmic Ecosystem Resolver in 3D (hereafter MACER3D) project, a three-dimensional hydrodynamic framework representing a comprehensive upgrade of MACER2D. MACER3D incorporates numerous improvements in physical processes and subgrid models, particularly focusing on gas and stellar feedback physics, including cooling/heating processes, SN feedback, and metal yielding. The framework enables controlled investigation of physical complexities while isolating AGN feedback effects from other feedback mechanisms.

The paper is organized as follows. In \S\ref{sec:methods}, we introduce the updated physical processes and setup in the MACER3D framework. In \S\ref{sec:results}, we present the results from the simulations of an isolated elliptical galaxy under the framework. Finally, we discuss and summarize our main conclusions in \S\ref{sec:discussion}.

\section{Methods} \label{sec:methods}

MACER3D incorporates a comprehensive upgrade of the previous MACER2D framework. Beyond the expansion in dimensionality, MACER3D incorporates a number of enhanced implementations of physical processes and subgrid models. Notable improvements primarily focus on gas and stellar feedback physics, including advanced cooling and heating processes, realistic SN feedback mechanisms, and detailed metal yielding. These enhancements are detailed in subsequent sections. The development roadmap of MACER3D includes the incorporation of non-thermal physics such as magnetic fields (Xia et al. in prep), and extends the framework to diverse galactic systems, including disk galaxies (Zou et al. in prep) and dwarf galaxies (Su et al. in prep), which will be presented in future work.

\subsection{Dimensionality and code base}

A fundamental enhancement in MACER3D is the implementation of three-dimensional physics, enabling more realistic modeling of AGN feedback effects on galaxy evolution. This advancement offers two key benefits. First, it allows for accurate representation of turbulence, which is crucial for understanding multiphase gas and metal mixing in the ISM and CGM. This represents a significant improvement over two-dimensional simulations, which suffer from inverse energy cascade, leading to artificial large-scale eddy formation and suppression of small-scale turbulence and mixing \citep{fjortoft53}. Second, the three-dimensional framework enables the simulation of non-axisymmetric structures and instabilities, including spiral arms, gravitational torques, and thermal instabilities, which significantly influence galaxy evolution and gas accretion onto the central supermassive black hole (SMBH) \citep{balbus89, binney}.

The numerical foundation of MACER3D has been modernized through the adoption of the Athena++ code \citep{white16}, replacing the ZEUS code \citep{stone92, hayes06} used in MACER2D. Athena++ is a state-of-the-art, open-source, high-performance, grid-based hydrodynamic code optimized for astrophysical applications. Its implementation of the Godunov method with directionally unsplit and staggered-mesh (USM) schemes ensures robust solution of hydrodynamic equations. The code's efficient CPU parallelization and its GPU-compatible version, AthenaK \citep{stone24}, provide opportunities for future computational enhancements.

The simulation solves three-dimensional hydrodynamic equations in spherical coordinates ($r$, $\theta$, $\phi$) using the standard Euler form, incorporating specialized gas physics and feedback mechanisms detailed in subsequent sections. The computational domain spans from $r_\mathrm{in} = 25 \,\mathrm{pc}$ to $r_\mathrm{out} = 250\,\mathrm{kpc}$, enabling simultaneous resolution of gas accretion flows across the SMBH Bondi radius \citep{bondi44} and large-scale galactic and halo environmental effects. The fiducial resolution is $256\times64\times128$, where radial grid spacing decreases logarithmically from outer to inner boundaries, achieving sub-pc resolution at the inner boundary. While this sub-pc resolution substantially increases computational demands for $\mathrm{Gyr}$ timescale evolution, it remains essential for accurate capture of accretion flows and AGN feedback processes. Under this configuration, the number of total timestep cycles for the simulation is a few $10^7$ to reach the total duration of $1.3$ Gyr for the fiducial simulation (and $1$ Gyr for other simulations with varying feedback prescriptions). To maintain computational efficiency, the simulation excludes a $6^\circ$ region near the z-axis to avoid geometric singularities.

\subsection{Initial and boundary conditions} \label{sec:ic}

The initial conditions of our simulations consist of a supermassive black hole (SMBH) of mass $M_\mathrm{bh}$ at the center of a dark matter halo (DM) which is modeled as a spherically symmetric quasi-isothermal halo with a circular velocity $v_\mathrm{c}$, and embedded within a stellar distribution and gas distribution in hydrostatic equilibrium. The stellar component follows the Jaffe profile \citep{jaffe83}:
\begin{equation}
    \rho_*=\frac{M_*r_*}{4\pi r^2(r_*+r)^2},
\end{equation}
where $M_*$ represents the total stellar mass and $r_*$ denotes the galaxy scale length. The distribution of gas number density $n_\mathrm{g}$ is characterized by the beta model \citep{mo10}:
\begin{equation}
    n_\mathrm{g}=n_0\left(1+\frac{r^2}{r_c^2}\right)^{-\frac{3\beta}{2}},
\end{equation}
where $n_0$ is the central gas number density and $r_c$ is the core radius and the beta parameter $\beta = 2/3$. The gas is initialized without rotation and configured to maintain hydrostatic equilibrium within a dark matter halo. The total mass profile is designed to follow an $r^{-2}$ law, consistent with observational constraints \citep{czoske08, dye08}. The gas metallicity is initially set to $2 Z_\odot$ within $0.125 r_c$ and decreases following a power-law profile with radius, reaching sub-solar values beyond $r_c$. We also note that dark matter halo and stellar component are static in our simulations, therefore the dynamical response of the gravitational potential to gas inflows, outflows, and feedback processes is not captured. We suspect that in the case of an elliptical galaxy, where the stellar component is typically dispersion-supported and evolves more slowly compared to disk galaxies, the impact of this approximation may be less severe. However, we acknowledge that this is a limitation of our current model and will be addressed in future work.

For boundary conditions, the outer boundary employs a modified outflow boundary condition: the standard outflow boundary condition allows the gas to enter or leave the computational domain across the outer boundary with zero gradient for density and velocity, while a non-zero pressure gradient with $\partial_r P = \rho g$ is always enforced at the outer boundary in order to preserve the hydrostatic equilibrium of the gas, where $g$ is the local gravitational acceleration measured at the outer boundary. At the inner boundary, we employ conditions that permit gas to leave the computational domain which is treated as black hole accretion, and simultaneously inject mass and momentum into the computational domain at certain rates and opening angles to simulate the AGN feedback-driven outflows, with detailed prescriptions described in \S\ref{sec:agn_fb}.

The parameters of the initial conditions are summarized in Table~\ref{tab:ic}. The initial conditions are set to match those of the fiducial MACER2D model as closely as possible for comparison purposes. The only significant difference is the initial central gas density is much higher in MACER3D, consistent with observations of ellipticals (e.g., \citealt{capelo2010hydrostatic,werner2012thermodynamic}), while the initial gas density in MACER2D is negligible and relies on stellar winds for gas supply.

\begin{table}
    \centering
    \begin{tabular}{cc}
        \hline
        Parameter & Value \\
        \hline
        DM halo circular velocity, $v_\mathrm{c}$ & $400\,\mathrm{km\ s^{-1}}$ \\
        SMBH mass, $M_\mathrm{bh}$ & $1.8\times 10^9\,\mathrm{M_\odot}$ \\
        Stellar mass, $M_*$ & $3\times 10^{11}\,\mathrm{M_\odot}$ \\
        Stellar scale length, $r_*$ & $9.27\,\mathrm{kpc}$ \\
        Central gas number density, $n_0$ & $0.08\,\mathrm{cm^{-3}}$ \\
        Gas core radius, $r_c$ & $6.9\,\mathrm{kpc}$ \\
        \hline
    \end{tabular}
    \caption{Key parameters used for initial conditions.}
    \label{tab:ic}
\end{table}

\subsection{Two-mode AGN feedback physics} \label{sec:agn_fb}

MACER3D implements AGN feedback through a sophisticated subgrid model based on its predecessor MACER2D \citep{yuan18}. The model distinguishes between two primary feedback modes: the hot (radio) mode and the cold (quasar) mode, delineated by a critical BH accretion rate $\dot{M}_\mathrm{BH} \sim 0.02\dot{M}_\mathrm{Edd}$ \citep{yuan14}. The hot mode occurs when low-density gas accretes onto the central SMBH, driving radio-mode feedback, while the cold mode activates during high-density gas accretion, powering quasar-mode feedback. Although the hot mode encompasses winds, jets, and radiation, we defer the implementation of jet feedback to future work for simplicity. For the cold mode, we consider only winds and radiation, excluding jets -- a choice consistent with observations showing that radio-loud quasars constitute only a small fraction of the total quasar population \citep{Kellermann1989,Ivezic2002,Banados2015,liu2021}.

The AGN feedback mode, hot or cold, is determined by the accretion rate onto the central SMBH $\dot{M}_\mathrm{BH}$, which is calculated on the fly. Since the inner boundary of our simulation is set at $r_\mathrm{in} = 25\,\mathrm{pc}$, which is typically smaller than the Bondi radius, we can directly compute the accretion rate crossing the inner boundary $\dot{M}(r_\mathrm{in})$, from which the BH accretion rate $\dot{M}_\mathrm{BH}$ and the properties of wind (outflow) can be inferred via the standard black hole accretion theory and observations, as detailed below. 

\subsubsection{The hot (radio) mode}

The accretion flow in the hot AGN mode is modeled to consist of two distinct regions: a truncated thin disk at large radii and a hot accretion flow within the truncation radius \citep{yuan14}. The truncation radius is given by:
\begin{equation}
    r_\mathrm{tr}\approx 3 r_\mathrm{s} \left[\frac{0.02\,\dot{M}_\mathrm{Edd}}{\dot{M}(r_\mathrm{in})}\right]^2
\end{equation}
where $r_\mathrm{s}$ denotes the Schwarzschild radius. While observational evidence for winds from hot accretion flows  \citep{wang13,cheung16,peng20,shi21,2022ApJ...926..209S} and its interaction with ISM \citep{2024ApJ...970...48S} has emerged recently, robust constraints on their properties remain limited. Therefore, following \citet{yuan18}, we adopt the theoretical prescriptions from \citet{yuan15}:
\begin{align}
    \dot{M}_\mathrm{BH} &= \dot{M}(r_\mathrm{in})\left(\frac{3r_\mathrm{s}}{r_\mathrm{tr}}\right)^{0.5}, \\
    \dot{M}_\mathrm{wind, hot} &= \dot{M}(r_\mathrm{in})-\dot{M}_\mathrm{BH}, \\
    v_\mathrm{wind\, hot} &= (0.2-0.4)v_\mathrm{K}(r_\mathrm{tr}),
\end{align}
where $v_\mathrm{K}(r_\mathrm{tr})$ represents the Keplerian velocity at the truncation radius. Consistent with \citep{yuan18}, we restrict the angular distribution of the wind to $30^\circ-70^\circ$ and $110^\circ-150^\circ$.

For radiative processes, we implement the detailed calculations of hot accretion flow radiative efficiency from \citet{xie12}. This efficiency is significantly lower than that of standard thin disks and exhibits strong dependence on the accretion rate:
\begin{equation}
    \varepsilon_\mathrm{hot}(\dot{M}_\mathrm{BH})=\varepsilon_0\left(\frac{\dot{M}_\mathrm{BH}}{0.01\dot{M}_\mathrm{Edd}}\right)^a=\varepsilon_0\left(100\,\dot{m}\right)^a,
\end{equation}
where the parameters $\varepsilon_0$ and $a$ vary across different accretion rate regimes:
\begin{equation}
    (\varepsilon_0,a)=\left\{
        \begin{aligned}
           &(0.2, 0.59), &\dot{m}\lesssim 9.4\times 10^{-5}\\
           &(0.045, 0.27), &9.4\times 10^{-5}\lesssim\dot{m}\lesssim 5\times 10^{-3}\\
           &(0.88, 4.53), &5\times 10^{-3}\lesssim\dot{m}\lesssim 6.6\times 10^{-3}\\
           &(0.1, 0), &6.6\times 10^{-3}\lesssim\dot{m}\lesssim 2\times 10^{-2}
        \end{aligned}
        \right.
\end{equation}

\subsubsection{The cold (quasar) mode}

When the black hole accretion rate exceeds the 2\% Eddington accretion rate, the accretion enters the cold mode. The cold mode is further divided into two regimes, bounded by the Eddington rate. Below the Eddington rate, the accretion flow is described by the standard thin disk. 10\% of the accretion power is converted to radiation, producing a highly luminous AGN with bolometric luminosity $L_\mathrm{bol}=0.1\dot{M}_\mathrm{BH}c^2$. The readers are referred to \citet{yuan18} for the calculation of $\dot{M}_\mathrm{BH}$. Leveraging extensive observational constraints on outflows in luminous AGN, we adopt the empirically fitted relations for mass flux and velocity as functions of $L_\mathrm{bol}$ from \citet{gofford15}, consistent with \citet{yuan18}:
\begin{equation}
    \dot{M}_\mathrm{wind,cold}=0.28\left(\frac{L_\mathrm{bol}}{10^{45}\,\mathrm{erg\, s^{-1}}}\right)^{0.85} \mathrm{{M_\odot}\, yr^{-1}},
\end{equation}
\begin{equation}
    v_\mathrm{wind,cold}=\min\left(2.5\times10^4\left(\frac{L_\mathrm{bol}}{10^{45}\,\mathrm{erg\, s^{-1}}}\right)^{0.4}, 10^5\right)\mathrm{km\, s^{-1}}.
\end{equation}
We impose an upper limit on $v_\mathrm{wind,cold}$ to reflect the observed velocity saturation \citep{gofford15}. Although small-scale accretion disk simulations suggest that outflows predominantly emerge at the spherical polar angles $\theta$ between $0^\circ-60^\circ$ and $120^\circ-180^\circ$ \citep{wang22}, we adopt a $\dot{M}(\theta)\approx \cos^2\theta$ distribution since the outflow opening angle significantly expands as it propagates outward to our inner boundary at $\sim 10^5 r_\mathrm{s}$.

For super-Eddington accretion, similar to the case of hot accretion mode, we do not have abundant constraint on the properties of wind. Therefore the wind mass flux and velocity in our model are taken from the theoretical study of wind and jet  based on general relativity radiative transfer MHD simulations of super-Eddington accretion flows around spinning black holes \citep{yang23}:
\begin{align}
    \dot{M}_\mathrm{BH} &= \left(\frac{3r_\mathrm{s}}{r_\mathrm{d}}\right)^{0.39}\dot{M}(r_\mathrm{d}), \\
   \dot{M}_\mathrm{wind,super} &= \dot{M}(r_\mathrm{d}) - \dot{M}_\mathrm{BH}, \\
    v_\mathrm{wind,super} &\approx 0.15 c,
\end{align}
where $r_\mathrm{d}$ is the outer boundary of the super-Eddington accretion disk with $r_\mathrm{d}= \min(r_\mathrm{circ},r_\mathrm{in})$, $r_\mathrm{circ}$ is the circularization radius of the accretion flow, and $\dot{M}(r_\mathrm{d}) \approx \dot{M}(r_\mathrm{in})$. The mass flux is confined to the spherical polar angle $\theta$ within $0^\circ-30^\circ$ and $150^\circ-180^\circ$. Following \citet{zhu23a}, we implement a radiative efficiency model fitted from \citet{jiang19}:
\begin{equation}
    \varepsilon_\mathrm{super}=0.21\left(\frac{100\dot{M}_\mathrm{BH}}{\dot{M}_\mathrm{Edd}}\right)^{-0.17}.
\end{equation}

\subsection{Star formation and stellar evolution} \label{sec:sf}

The star formation and stellar evolution subgrid model follows the prescriptions in MACER2D \citep{ciotti12}. We briefly summarize the key aspects here. The implementation is based on the Schmidt-Kennicutt law \citep{kennicutt98} and stellar evolution models \citep{maraston05}. Gas that exceeds a critical number density threshold $n_\mathrm{th} = 1\,\mathrm{cm^{-3}}$ and falls below a critical temperature threshold $T_\mathrm{th} = 10^{4}\,\mathrm{K}$ undergoes conversion to stars. The conversion rate depends on a star formation efficiency parameter $\epsilon_\mathrm{SF} = 0.1$ and a star formation timescale $\tau_\mathrm{SF}$:
\begin{equation}
\tau_\mathrm{SF}= \max(\tau_\mathrm{cool},\tau_\mathrm{dyn}),
\end{equation}
where $\tau_\mathrm{cool}$ is the cooling timescale computed via the cooling algorithm (detailed in \S\ref{sec:coolheat}), and $\tau_\mathrm{dyn}$ is the dynamical timescale determined by the minimum of the local free-fall time and rotational timescale:
\begin{equation}
\tau_\mathrm{dyn} = \min(\tau_\mathrm{ff}, \tau_\mathrm{rot}),
\end{equation}
where
\begin{align}
\tau_\mathrm{ff} &\equiv \sqrt{\frac{3\pi}{32G\rho}}, \\
\tau_\mathrm{rot} &\equiv \sqrt{\frac{r \partial \Phi(r)}{\partial r}},
\end{align}
with $\Phi(r)$ being the gravitational potential. The star formation rate is then given by:
\begin{equation}
    \dot{M}_\mathrm{SF}=\frac{\epsilon_\mathrm{SF}\,\rho}{\tau_\mathrm{SF}}.
\end{equation}
The model evolution is primarily driven by stellar mass loss and SNe Ia rates associated with the initial stellar distribution. Following \citep{ciotti12}, we express the stellar mass loss as a piece-wise function:
\begin{equation}
    \Delta M=\begin{cases}
    0.945M_\mathrm{TO}-0.503\mathrm{M_\odot}, &\text{if}\ M_\mathrm{TO}<9\mathrm{M_\odot}\\
    M_\mathrm{TO}-1.4\mathrm{M_\odot}, &\text{if}\ M_\mathrm{TO}\geq9\mathrm{M_\odot}
    \end{cases},
\end{equation}
where the stellar turn-off mass $M_\mathrm{TO}$ follows the relation:
\begin{equation}
    \log\frac{M_\mathrm{TO}}{\mathrm{M_\odot}}=0.0558\left(\log\frac{t}{1\,\mathrm{yr}}\right)^2-1.338\log\frac{t}{1\,\mathrm{yr}}+7.764.
\end{equation}

We note that while star formation and evolution models in the literature exhibit considerable variation in parameters and criteria, we have deliberately adopted a simplified prescription. This approach minimizes free parameters and maintains the idealized nature of the simulation, facilitating focused investigation of AGN feedback effects in a controlled environment.

\subsection{Supernova feedback physics} \label{sec:sn_fb}

In MACER2D, supernova feedback was implemented as thermal energy injection proportional to the local SN event rate. Without identifying individual SN events, the energy injection was spatially smoothed, acting as an extra heating term in the energy equation of the hydrodynamic simulation. While computationally efficient, this simplified approach did not accurately capture the effects of discrete supernova explosions, such as shock propagation and turbulence generation in the ISM and CGM driven by individual events. MACER3D significantly improves upon this by implementing a more sophisticated model of supernova feedback with state-of-the-art prescriptions.

First, we model supernovae as discrete events. Given their independent nature, we use a Poisson distribution to determine the probability of SN events occurring:
\begin{equation}
    P(N_\mathrm{SN}; \mu_\mathrm{SN})=\frac{e^{-\mu_\mathrm{SN}}\mu_\mathrm{SN}^{N_\mathrm{SN}}}{N_\mathrm{SN}!},
\end{equation}
where $\mu_\mathrm{SN}$ is the expected number of SN events per unit time in a given region, and $P(N_\mathrm{SN}; \mu_\mathrm{SN})$ gives the probability of observing $N_\mathrm{SN}$ events. The expectation value $\mu_\mathrm{SN}$ in a given time step $\Delta t$ is determined by the sum of the volume-integrated type Ia and type II SNe rates:
\begin{equation}
    \mu_\mathrm{SN}= (R_\mathrm{II} + R_\mathrm{Ia})\,\Delta t,
\end{equation}
where $R_\mathrm{II}$ and $R_\mathrm{Ia}$ are the type II and type Ia SN rates, respectively, which are adopted from \citet{ciotti12}:
\begin{equation}
    R_\mathrm{II}=\frac{\epsilon_\mathrm{II}}{\tau_\mathrm{II}M_\mathrm{II,ZAMS}}\int_0^t\dot{M}_\mathrm{SF} (t')e^{-\frac{t-t'}{\tau_\mathrm{II}}}\mathrm{d}t',
\end{equation}
where $\epsilon_{\mathrm{II}} = 0.1234$ is the type II SN efficiency, reflecting the fraction of stellar mass from $9$ -- $120\,\mathrm{M_\odot}$ progenitors undergoing core collapse, $\tau_\mathrm{II} = 2 \times 10^7\,\mathrm{yr}$ the characteristic type II SN timescale, and $M_\mathrm{II,ZAMS} = 21.34\,\mathrm{M_\odot}$ the IMF-averaged ZAMS mass of type II SN progenitors. The type Ia SN rate is given by:
\begin{equation}
    R_\text{Ia}(t)=R_0h^2\frac{L_\text{B}}{L_{\text{B}\odot}}\left(\frac{t}{13.7\,\mathrm{Gyr}}\right)^{-s}\,\mathrm{yr^{-1}},
\end{equation}
where $R_0 = 0.22\times10^{-12}$ \citep{cappellaro99,maoz14}, slightly lower than \citet{ciotti12}, $h = H_0/100\,\mathrm{km\,s^{-1}\,Mpc^{-1}} = 0.75$ \citep{cappellaro99}, $L_\mathrm{B}$ is the B-band stellar luminosity, and $s=1.1$. At each time step $\Delta t$, we sample a local random number from this Poisson distribution to determine the number of SN events. A key property of the Poisson distribution ensures that the accumulated samples over time converge to the expectation value $\mu_\mathrm{sn}$, maintaining consistency with the prescribed SN rate in the simulation. Each type Ia SN injects $M_\mathrm{Ia} = 1.4\,\mathrm{M_\odot}$ into the ambient gas, while type II SNe inject an IMF-averaged mass of $M_\mathrm{II} = 16.6\,\mathrm{M_\odot}$, based on \citet{sukhbold16}. The total energy injected into the ISM per supernova is $E_\mathrm{SN} = 0.85 \times 10^{51}\,\mathrm{erg}$.

Another significant enhancement in MACER3D is the implementation of an improved supernova energy injection prescription. We implement the scaling relation developed by \citet{martizzi15}, derived from high-resolution simulations of supernova explosions in an inhomogeneous medium with varying densities and metallicities. This subgrid prescription offers two key advantages. First, it incorporates momentum feedback, enabling the simulation to address the overcooling problem by directly accounting for momentum transfer from supernovae. This is particularly crucial in high-density regions where, due to limited numerical resolution, radiative cooling would otherwise artificially dominate and suppress realistic feedback effects. Second, the prescription accurately captures both turbulent and thermal energy injection from multiple concurrent supernova events \citep{martizzi15}, which is essential for modeling clustered supernova explosions during periods of elevated supernova activity. While this prescription is typically enabled in MACER3D, particularly for disk (Zou et al. in prep) and dwarf galaxy simulations (Su et al. in prep), \citet{martizzi15} also mention the heads-up that their fitting formulas are not calibrated for and thus less accurate in low-density ambient medium due to the longer cooling time, therefore for the simulated ellipticals in this work which indeed have a low-density ISM, we adopt the thermal feedback that is widely used for simulations of ellipticals in the literature (e.g. \citealt{sharma2014hot,li2020impact}). We also note that the SN fade radius ranging between $20\,\mathrm{pc}$ to $150\,\mathrm{pc}$ raised by \citet{li2020impact}  is well resolved in our simulations, where the mean resolution within $r < 1\mathrm{kpc}$ (where most of the SN events occur) is $\sim 7\,\mathrm{pc}$.

\subsection{Radiative cooling and heating} \label{sec:coolheat}

Although star formation in MACER3D is implemented as a subgrid model that does not directly involve cooling for molecular cloud formation, accurate treatment of radiative cooling remains critical for calculating the cool gas supply to the central SMBH and star formation. Recent idealized, small-scale numerical studies have emphasized the fundamental role of radiative cooling in the formation, survival, and destruction of cool ($\sim 10^4\,\mathrm{K}$) gas in the ISM and CGM \citep{armillotta2016efficiency,armillotta2017survival,gronke18,gronke2022survival} through thermal instabilities \citep{mccourt12,sharma12,ji18} and turbulent mixing \citep{ji19,fielding20,Tan21,yang2023radiative}. With these considerations in mind, we have significantly enhanced the radiative cooling and heating model in MACER3D.

\begin{figure}
    \centering
    \includegraphics[width=\linewidth]{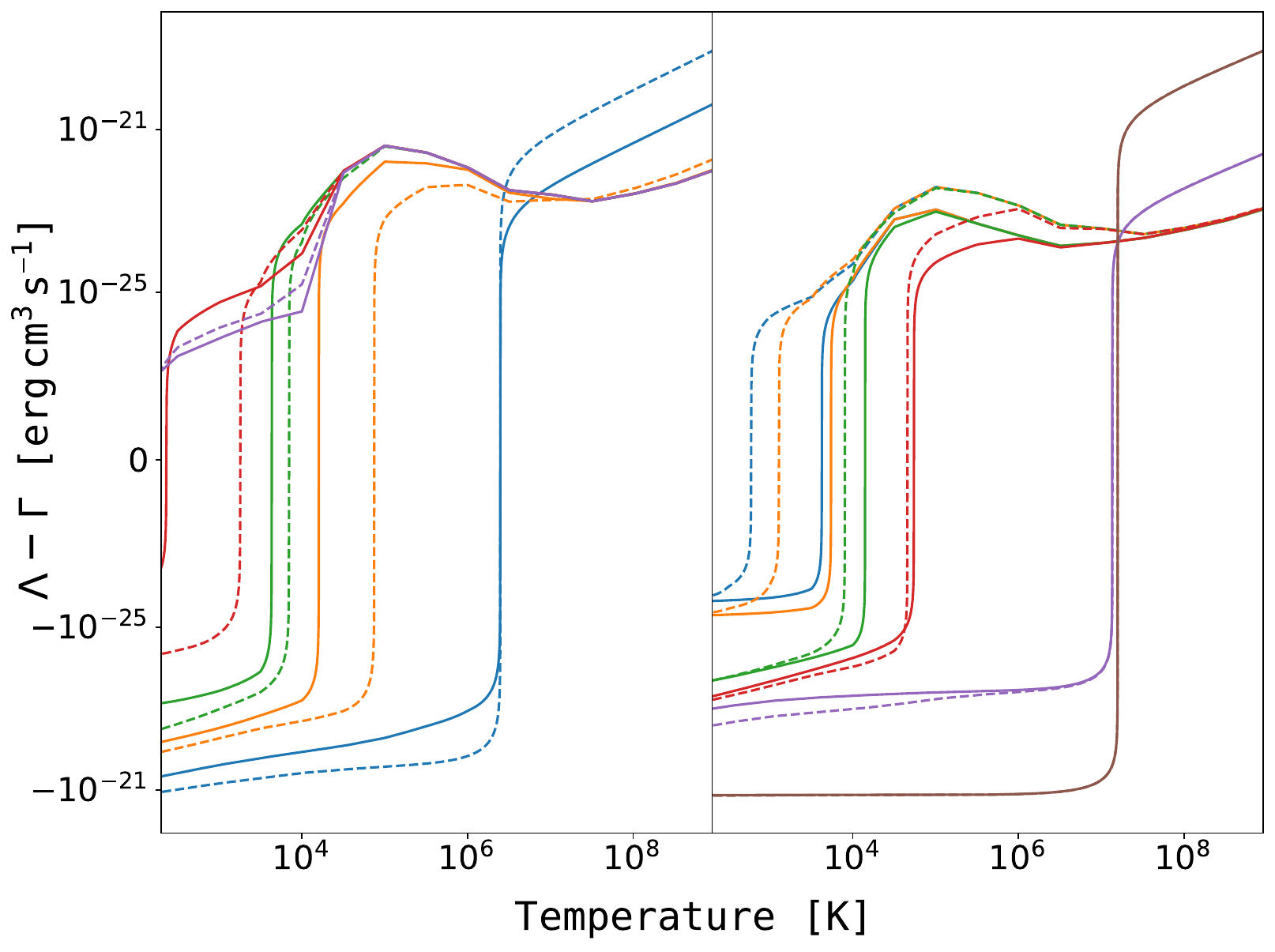}
    \caption{A representative subset of the cooling functions adopted in MACER3D, illustrating their dependence on gas number density (left) and AGN radiation flux (right). In the left panel, colors represent different number densities $n_\mathrm{H}$: $10^{-8} \, \mathrm{cm^{-3}}$ (blue), $10^{-5} \, \mathrm{cm^{-3}}$ (orange), $10^{-2} \, \mathrm{cm^{-3}}$ (green), $10 \, \mathrm{cm^{-3}}$ (red), and $10^4 \, \mathrm{cm^{-3}}$ (purple). Solid and dashed lines correspond to redshifts of $z=0$ and $z=2$, respectively, with zero AGN flux and solar metallicity. In the right panel, colors represent varying AGN radiation flux: $0 \, \mathrm{erg \, s^{-1} \, cm^{-2}}$ (blue), $10^{-5} \, \mathrm{erg \, s^{-1} \, cm^{-2}}$ (orange), $10^{-2} \, \mathrm{erg \, s^{-1} \, cm^{-2}}$ (green), $10 \, \mathrm{erg \, s^{-1} \, cm^{-2}}$ (red), $10^4 \, \mathrm{erg \, s^{-1} \, cm^{-2}}$ (purple), and $10^7 \, \mathrm{erg \, s^{-1} \, cm^{-2}}$ (brown). Solid and dashed lines correspond to metallicities of $0.1 \, Z_\odot$ and $Z_\odot$, with a fixed number density of $1 \, \mathrm{cm^{-3}}$ and redshift of $z=0$.}
    \label{fig:coolingtable}
\end{figure}

The first enhancement involves implementing more accurate cooling functions. Different from the empirical fitting formula \citep{sazonov05} used in MACER2D, we employ a comprehensive 5-dimensional cooling table generated using Cloudy \citep{cloudy}. This table captures the dependence on gas density $n_\mathrm{H}$, temperature $T$, metallicity $Z$, redshift $z$, and AGN radiative flux $F_\mathrm{AGN}$. The redshift dependence incorporates both ultraviolet background (UVB) and cosmic microwave background (CMB) radiation. The cooling function spans an extensive range of physical parameters relevant to our simulations: $n_\mathrm{H}=10^{-8}-10^4\,\mathrm{cm^{-3}}$, $Z=10^{-10},\,10^{-3}-10\,Z_\odot$, $z=0-10$, $T=10^2-10^9\,\mathrm{K}$, and $F_\mathrm{AGN}=0,\,10^{-7}-10^7\,\mathrm{erg\,s^{-1}\,cm^{-2}}$. Figure~\ref{fig:coolingtable} illustrates a representative subset of the cooling functions, in particular, showcasing their dependence on gas density and AGN radiation flux.

The second enhancement in MACER3D is the implementation of the Townsend exact cooling integration scheme \citep{townsend}, which offers superior robustness and precision compared to conventional explicit or implicit time integration schemes. This approach reformulates the energy equation into an operator-split form:
\begin{equation}
    \int_{T^n}^{T^{n+1}}\frac{\mathrm {d}T}{\Lambda(T)}=-\frac{(\gamma-1)\mu\rho}{k_\mathrm{B}\mu_e\mu_im_\mathrm{p}}\Delta t,
\end{equation}
where $\gamma$ denotes the adiabatic index, $\mu$ the mean molecular weight, $\mu_i$ ($\mu_e$) the mean molecular weight per ion (electron), and $m_\mathrm{p}$ the proton mass, with $T^n$ and $T^{n+1}$ representing the temperatures at steps $n$ and $n+1$, respectively. The mean molecular weight is updated at each time step based on local metallicity, with the full ionization approximation since individual species are not traced in our simulations given both the computational cost and complexity, and the star formation is consequently treated as a subgrid model as described in \S\ref{sec:sf}. For piecewise power-law cooling functions $\Lambda(T)$, this operator-split equation permits analytical solutions through integration from a reference temperature $T_\mathrm{ref}$ to the current temperature $T^n$, enabling exact calculation of the new temperature $T^{n+1}$ for arbitrary time steps. This scheme facilitates more accurate and efficient temperature evolution by eliminating constraints from the Courant condition due to short cooling times, thereby mitigating overcooling issues. We also note that no time step limit constrained by the cooling time is imposed in our work, which is allowed by the Townsend scheme, however, the time step limit from the Courant condition in our simulations is typically as short as hundreds of years due to the finest resolution, which is sufficiently small compared to the cooling time.

We further extend the Townsend cooling scheme to incorporate heating processes, including photoionization and Compton heating from AGN and UVB radiation. This extension introduces equilibrium points in the cooling curves where heating balances cooling (indicated by net cooling curves crossing zero on the y-axis in Fig.~\ref{fig:coolingtable}), which require special consideration in the integration scheme.\footnote{The dimensionless temporal evolution function in Eq. (24) of \citet{townsend}, which describes temperature evolution by integrating the cooling function from an arbitrary reference temperature, approaches infinity at the equilibrium points by definition. However, this singularity is purely mathematical, and is eliminated by choosing two different reference temperatures and performing integrations on each side of the equilibrium point along the heating and cooling branches, respectively.} The modified scheme ensures accurate and simultaneous treatment of both cooling and heating processes while maintaining self-consistency in temperature evolution.

\subsection{Metallicity} \label{sec:metal}

Metallicity plays a fundamental role in many astrophysical processes, particularly in radiative cooling. Metals, synthesized through stellar nucleosynthesis, are injected into the ISM and CGM through SN explosions and stellar winds, and subsequently redistributed by turbulent diffusion. Therefore, incorporating metallicity evolution is essential for accurate simulation of galaxy evolution. For computational efficiency while maintaining physical accuracy, we track the evolution of total metallicity rather than individual elemental abundances, as this approach sufficiently captures the key physics, especially for cooling and heating calculations. Although implementing a more detailed metallicity model that tracks individual elements (e.g., \citealt{eisenreich2017active}) is valuable, such implementation is straightforward within the MACER3D framework and will be considered in future work.

The evolution of gas metallicity $Z_\mathrm{gas}$ is modeled as a passive scalar governed by sink terms from star formation and source terms from stellar yields, following the equation:
\begin{equation}
    Z_{\mathrm{gas}} \dot{\rho}_{\mathrm{gas}} = Z_{\mathrm{II}} \dot{\rho}_{\mathrm{II}} + Z_{\mathrm{Ia}} \dot{\rho}_{\mathrm{Ia}} + Z_{\mathrm{SE}} \dot{\rho}_{\mathrm{SE}} - Z_{\mathrm{gas}} \dot{\rho}_{\mathrm{SF}}, \label{eq:MetalEV}
\end{equation} 
where $\dot{\rho}_{\mathrm{II}}$, $\dot{\rho}_{\mathrm{Ia}}$, and $\dot{\rho}_{\mathrm{SE}}$ denote the mass loss rates from SN II, SN Ia, and stellar evolution respectively, while $\dot{\rho}_{\mathrm{SF}}$ represents the star formation rate. The corresponding metal yields are given by $Z_{\mathrm{II}}$, $Z_{\mathrm{Ia}}$, and $Z_{\mathrm{SE}}$. Although metal production from stellar winds is not explicitly included here, these contributions are incorporated into the stellar evolution models. While our implementation includes metal diffusion to account for additional mixing processes, this mechanism is deactivated in this initial study using the MACER3D framework.

For SN yields, we adopt values from \citet{Hopkins2018FIRE2}, with $Z_{\mathrm{Ia}} = 1$ for SN Ia and $Z_{\mathrm{II}} = 1.02(1.9134 + 0.0479 \Tilde{N})/10.5$ for SN II, where $\Tilde{N} = \max(Z_*/Z_{\odot},\ 1.65)$. These yields represent averages over the initial mass function. Given our simulation's initial redshift of $z \sim 2$, corresponding to a turn-off mass $M_{\mathrm{TO}} \sim 1.4\, \mathrm{M_\odot}$, metal production through stellar evolution primarily originates from low-mass stars ($M_* < 8 \mathrm{M_\odot}$) via planetary nebulae and stellar winds. To determine stellar evolution yields, we employ the yield tables from \citet{Nomoto2013YieldsTable}, implementing bilinear interpolation to construct a fitting function $Z_{\mathrm{SE}}$ dependent on both $M_{\mathrm{TO}}$ and stellar metallicity ($Z_*$).

\subsection{Simulation suit}

As the very first work of the MACER3D framework, we focus on the evolution of an isolated elliptical galaxy and explore the impact of the AGN feedback and SN feedback on the galaxy's evolution, respectively. We set up three simulations to investigate the effects of different feedback mechanisms on the galaxy's evolution. The {\tt fiducial} simulation includes both AGN feedback and SN feedback. Another two simulations, {\tt noAGNfb} and {\tt noSNfb}, turn off AGN feedback and SN feedback, respectively, while other feedback mechanisms (e.g., stellar wind) remain active. We note that although the AGN feedback is disabled in {\tt noAGNfb} simulation, a central SMBH is still included, therefore the central BH still accretes gas, grows and give rise to luminosity, but neither outflows nor heating are produced. In all simulations, multiple passive scalar tracers are included to track the evolution of the mass from different sources, including AGN winds (hot and cold traced separately), stellar winds, ejecta of SNe Ia and SNe II, and the ISM/CGM gas.

\section{Results} \label{sec:results}

\subsection{Time evolution of AGN luminosity and star formation rate}

\begin{figure*}
    \centering
    \includegraphics[width=\textwidth]{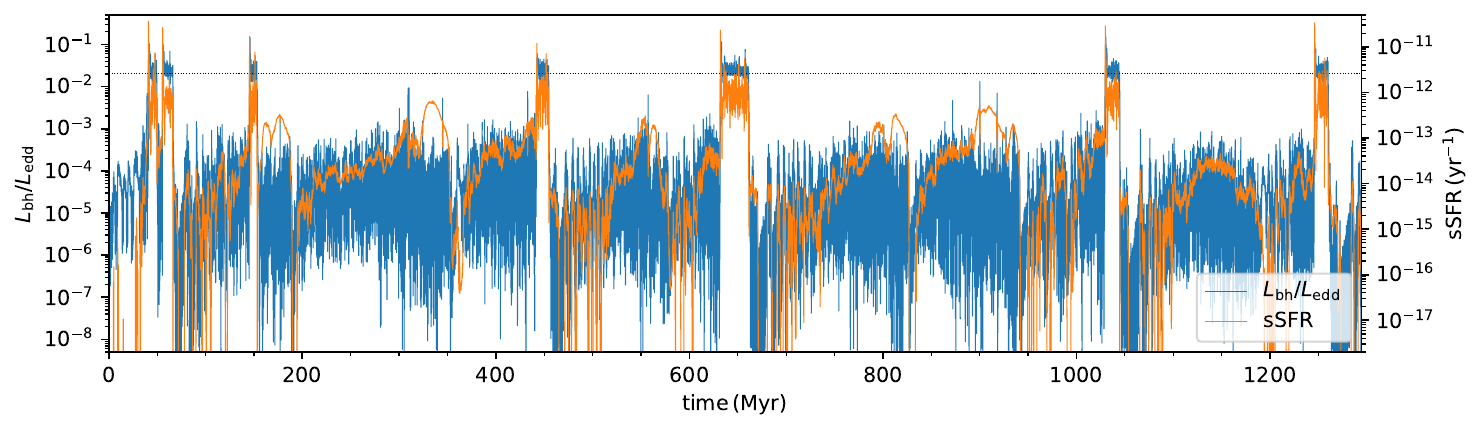}
    \caption{Time evolution of the AGN luminosity (blue, scaled on the left y-axis) and specific star formation rate (orange, scaled on the right y-axis), superposed with $L_\mathrm{bh} / L_\mathrm{Edd} = 2\%$ (dashed line), above (below) which the AGN feedback enters the cold (hot) mode. Both quantities fluctuate over time by orders of magnitude, with occasional bursts of activity when the AGN feedback enters the cold mode ($L_\mathrm{BH} \geq 2\times 10^{-2} L_\mathrm{Edd}$). The AGN luminosity and the specific star formation rate exhibit strong positive correlation temporally.}
    \label{fig:lum_sfr}
\end{figure*}

Fig.~\ref{fig:lum_sfr} presents the temporal evolution of the AGN luminosity $L_\mathrm{BH}$ (normalized by the Eddington luminosity $L_\mathrm{Edd}$) and the specific star formation rate (sSFR) in our fiducial simulation. Both quantities demonstrate pronounced temporal variability spanning multiple orders of magnitude. The AGN luminosity predominantly maintains a relatively low level of $L_\mathrm{BH}/L_\mathrm{Edd} \sim 10^{-5}$, punctuated by episodic bursts where $L_\mathrm{BH}/ L_\mathrm{Edd}$ exceeds $10^{-2}$, triggering cold-mode AGN feedback.

The sSFR evolution exhibits two distinct states while remaining consistently below $10^{-12}\,\mathrm{yr^{-1}}$, indicating the galaxy's quiescent nature. During the ``low state'', associated with hot-mode AGN feedback, the sSFR demonstrates a systematic increase from $10^{-16}\,\mathrm{yr^{-1}}$ to $10^{-12}\,\mathrm{yr^{-1}}$ over approximately $100\,\mathrm{Myr}$ timescales, likely reflecting gradual cold gas accumulation. The ``high state'', coinciding with cold-mode AGN feedback, is characterized by elevated sSFR levels of several $10^{-12}\,\mathrm{yr^{-1}}$, occasionally exceeding $10^{-11}\,\mathrm{yr^{-1}}$ at onset. These high-state episodes typically persist for approximately $10\,\mathrm{Myr}$ before rapidly declining to low-state values.

A striking feature is the strong temporal correlation between AGN luminosity and sSFR, suggesting shared dependence on available gas supply. Despite this positive correlation, evidence of AGN's negative feedback on star formation remains apparent: although sSFR initially spikes above $10^{-11}\,\mathrm{yr^{-1}}$ during high states, it promptly stabilizes at $\sim 10^{-12}\,\mathrm{yr^{-1}}$ under cold-mode AGN feedback, maintaining the galaxy's quiescent state. The sharp sSFR decline concluding each high state likely reflects cold gas depletion through intense AGN feedback. We note that although the powerful cold mode seems to dominate the star formation activity during the bursts, the hot mode is at least equally important in regulating the cool gas formation and suppressing the star formation over longer timescales. This will be further investigated in a separate work.

The AGN duty cycle manifests as periodic bursts of $L_\mathrm{BH}/L_\mathrm{Edd} \gtrsim 10^{-2}$ occurring at intervals of tens to hundreds of Myr, coincident with rapid central gas inflows. Notably, while our model permits super-Eddington accretion, no such events are observed throughout the simulation. These results collectively demonstrate the intricate coupling between AGN activity and star formation, which we examine in greater detail in subsequent sections.

\subsection{Spatial distribution of gas properties: at galactic halo scales}

\begin{figure*}
    \centering
    \includegraphics[width=\textwidth]{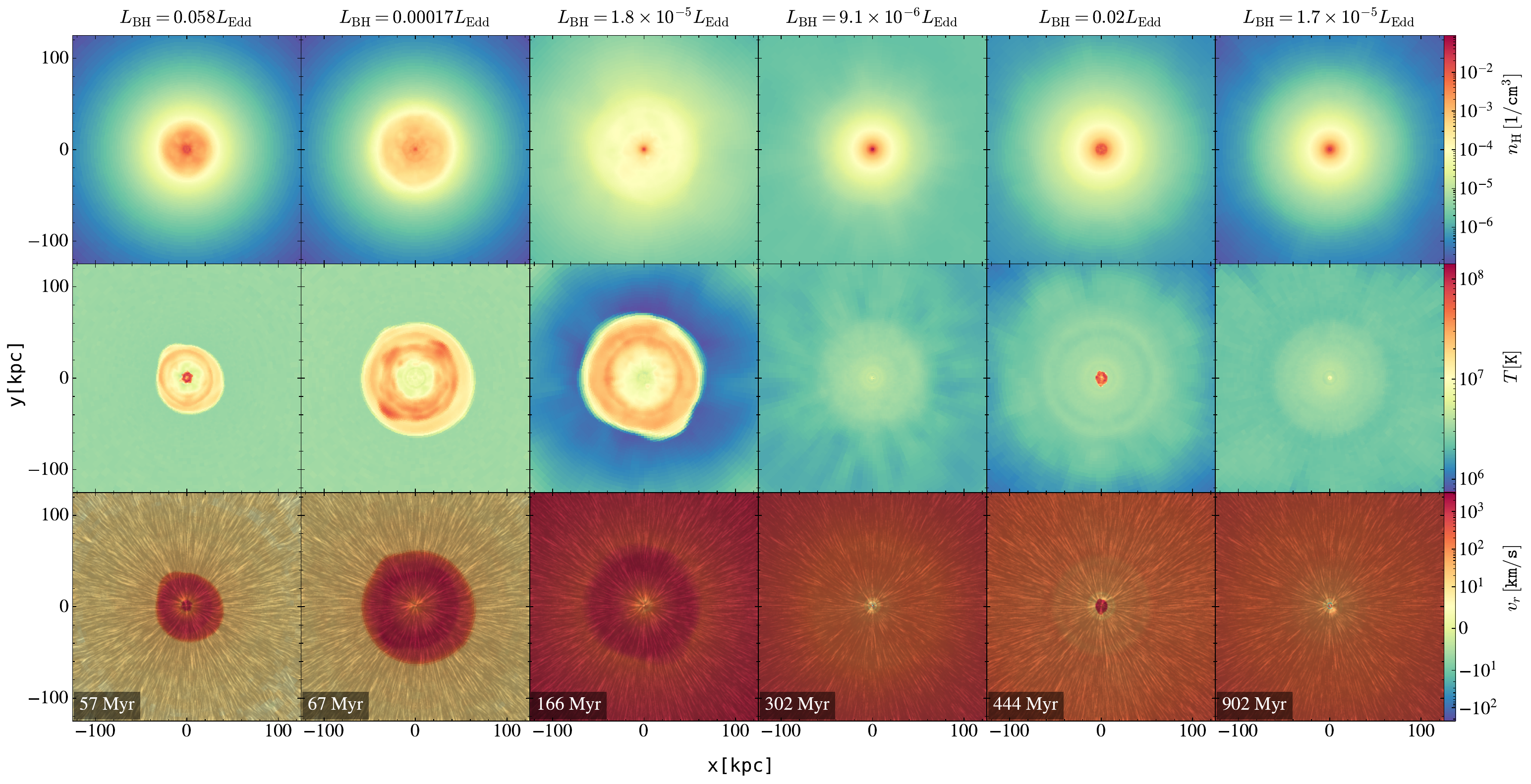}
    \caption{Projection plots of gas properties viewed from the polar angle within $100\,\mathrm{kpc}$ radius at different evolutionary stages with varying BH luminosity: volume-weighted number density $n_\mathrm{H}$ (top), mass-weighted temperature (middle), and mass-weighted radial velocity with overplotted velocity streamlines (bottom). The simulation time and AGN luminosities are annotated for each column. The halo-scale gas properties exhibit substantial temporal variations indicative of AGN-driven outflows and feedback effects, though without direct correlation to instantaneous AGN luminosity due to delayed response timescales of the halo gas.}
    \label{fig:theta_100}
\end{figure*}

We next examine the spatial distribution of key gas properties -- number density $n_\mathrm{H}$, temperature, and radial velocity -- at galactic halo scales. Fig.~\ref{fig:theta_100} presents polar-view projections of these quantities within a $100\,\mathrm{kpc}$ radius at different evolutionary stages characterized by varying AGN luminosities. The panels show volume-weighted number density (top), mass-weighted temperature (middle), and mass-weighted radial velocity with overplotted velocity streamlines (bottom).

The halo-scale gas properties demonstrate significant temporal evolution and spatial structure. Several characteristic features are evident:
\begin{enumerate}[label=(\roman*)]
    \item High-temperature, outward-expanding shells with pronounced density and temperature gradients, indicative of strong AGN-driven outflows ($t = 57$, $67$ and $166\,\mathrm{Myr}$);
    \item Central hot spots with $T \gtrsim 10^8\,\mathrm{K}$ signifying either recently launched AGN winds during high accretion rates or cold-mode AGN activity ($t = 57$ and $444\,\mathrm{Myr}$);
    \item Post-outflow regions characterized by intermediate temperatures and low densities, reflecting the aftermath of past AGN feedback episodes ($t = 302$ and $902\,\mathrm{Myr}$).
\end{enumerate}

Notably, no direct correlation exists between the instantaneous AGN luminosity and the halo-scale gas properties. This lack of immediate correspondence is consistent with the substantial difference between the halo gas dynamical timescale ($\sim 0.1\,\mathrm{Gyr}$) and the more rapid AGN variability timescale demonstrated in Fig.~\ref{fig:lum_sfr}. Consequently, the large-scale gas properties reflect the delayed and integrated effects of AGN feedback over extended periods rather than responding to instantaneous AGN activity. This temporal disconnect between central AGN behavior and halo-scale gas dynamics highlights the importance of considering different characteristic timescales when interpreting feedback effects across varying spatial scales.

\subsection{Spatial distribution of gas properties: from vicinity of Bondi radius to galactic scales} \label{sec:gas_small_scale}

\begin{figure*}
    \centering
    \includegraphics[width=\textwidth]{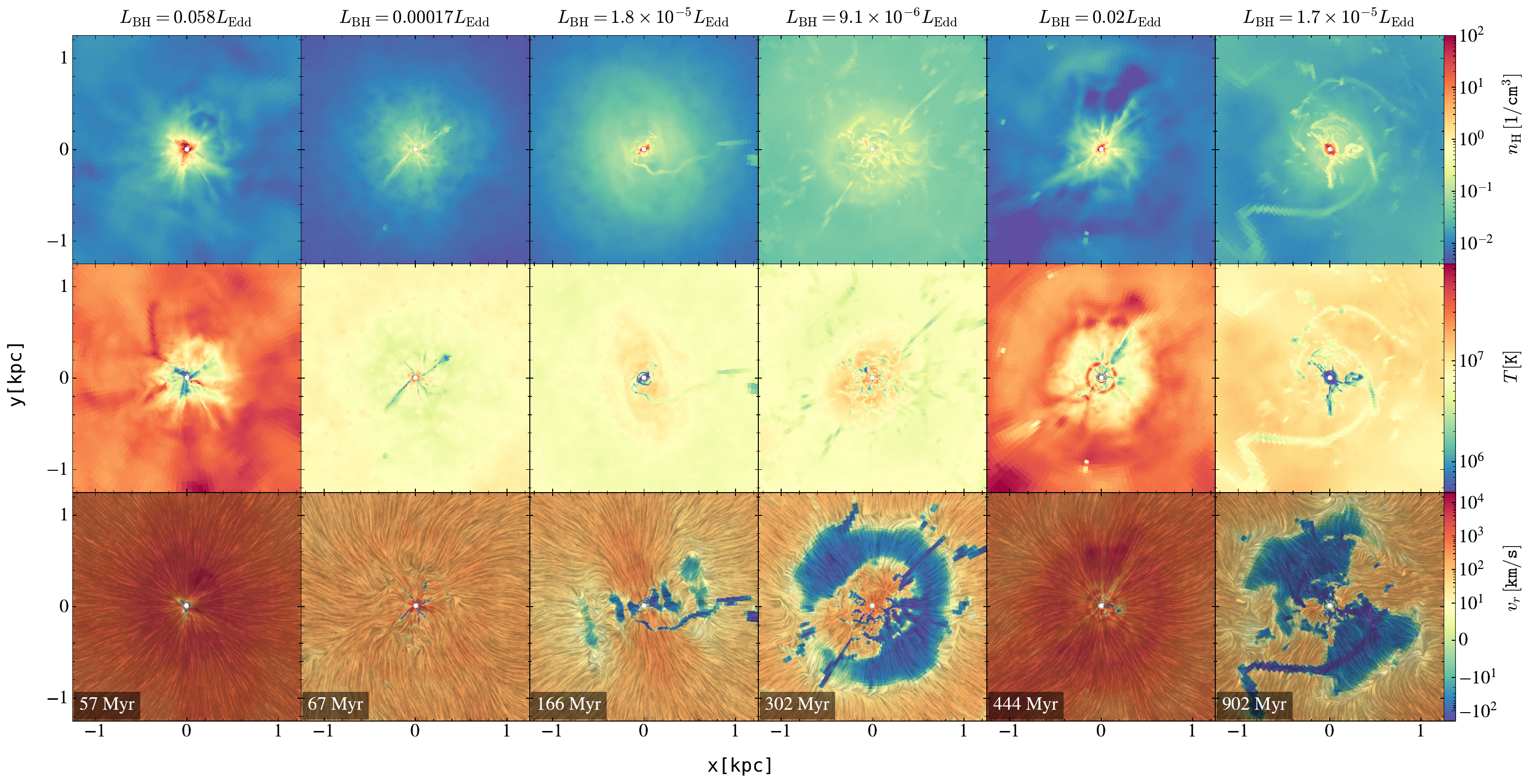}
    \caption{Projection plots as Fig.~\ref{fig:theta_100}, but zoomed in to a radius of $1\,\mathrm{kpc}$. At this scale, the gas properties exhibit more complex and detailed structures, and demonstrate strong correlation with the instantaneous AGN luminosity. At high accretion rates, high-density cool gas accretes onto the center, triggering high-velocity cold winds exceeding $10^4\,\mathrm{km/s}$ that shock-heat the ambient medium to temperatures approaching $10^8\,\mathrm{K}$. During low accretion periods, both gas densities and temperatures decrease significantly, and the gas motions become milder and more turbulent.}
    \label{fig:theta_1}
\end{figure*}

Fig.~\ref{fig:theta_1} follows the format of Fig.~\ref{fig:theta_100}, but focuses on the central kiloparsec where the Bondi radius (indicated by the white central dot) becomes visible. The gas properties at this scale reveal complicated structures that exhibit strong temporal correlation with AGN activity. During periods of high BH accretion ($t = 57\,\mathrm{Myr}$ and $444\,\mathrm{Myr}$), high-density cool gas forms at the vicinity of the central black hole via thermal instability, and accretes onto the center, triggering high-velocity cold winds exceeding $10^4\,\mathrm{km/s}$ that shock-heat the ambient medium to temperatures approaching $10^8\,\mathrm{K}$. Conversely, during low accretion periods, both gas densities and temperatures decrease significantly, indicating a more quiescent state. The velocity field demonstrates complex dynamics characterized by simultaneous inflows, outflows, and turbulent motions, and cool gas manifests as clumpy or spiral-structured filaments falling toward the central regions. These results highlight the close coupling between AGN feedback and gas dynamics near the Bondi radius, which critically determines both the black hole accretion rate and subsequent feedback modes. Notably, unlike the large-scale properties, gas conditions at $r \sim 1\,\mathrm{kpc}$ exhibit strong correlation with instantaneous AGN luminosity.

\begin{figure*}
    \centering
    \includegraphics[width=\textwidth]{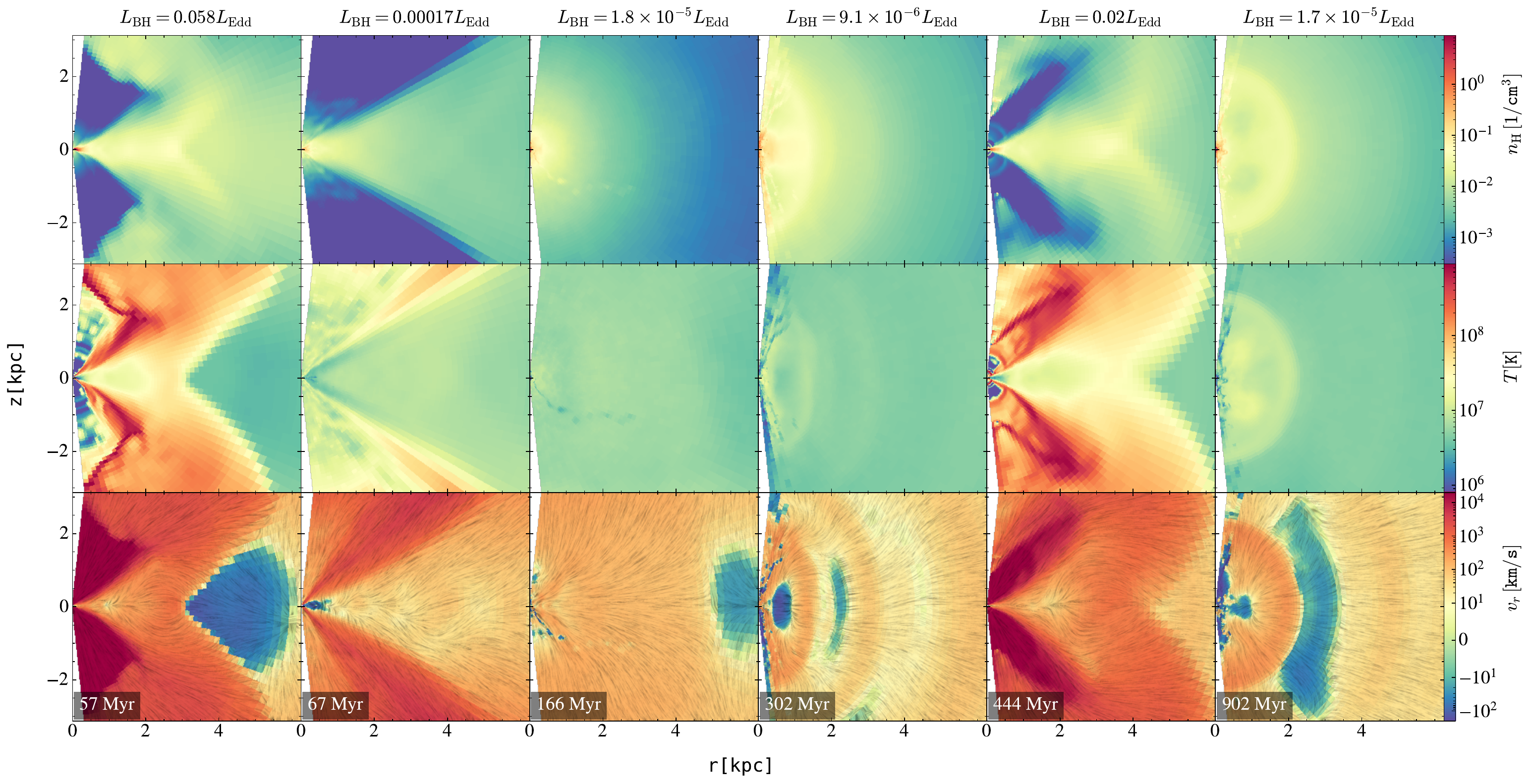}
    \caption{Projection plots as Fig.~\ref{fig:theta_100}, but viewed from the azimuthal angle and zoomed in to a radius of $5\,\mathrm{kpc}$. The feedback structures are more clearly visible from this perspective, revealing high-velocity ($>10^4\,\mathrm{km/s}$) cold winds in bi-conical regions during cold-mode AGN feedback, with prominent shock fronts at the wind-ISM interface.}
    \label{fig:phi_5}
\end{figure*}

Fig.~\ref{fig:phi_5} presents an azimuthal perspective within $5\,\mathrm{kpc}$ radius to reveal feedback structures. To identify the dominant feedback mechanisms, we look at the spatial distribution of the mass tracers (plots are not shown for the sake of brevity), which are passive scalars that are injected into the computational domain by the AGN cold winds, AGN hot winds, SN feedback, etc., respectively. By examining the spatial concentration of these tracers, we can identify the dominant feedback mechanism driving gas flows across different regions of the galaxy and at various times. During high accretion episodes ($t = 57\,\mathrm{Myr}$ and $444\,\mathrm{Myr}$), cold-mode AGN feedback drives high-velocity ($>10^4\,\mathrm{km/s}$), low-density winds in bi-conical regions, generating prominent shock fronts at the wind-ISM interface. During low accretion periods ($t = 166\,\mathrm{Myr}$, $302\,\mathrm{Myr}$, and $902\,\mathrm{Myr}$), hot AGN winds dominate the feedback. At $t = 67\,\mathrm{Myr}$, the transition from cold to hot mode feedback is evident: low-density bi-conical regions carved by previous cold winds persist while nascent hot winds begin launching. These observations demonstrate the complex, dynamic impact of AGN feedback on intermediate-scale gas properties, which plays a crucial role in regulating both gas dynamics and star formation of the galaxy.

\subsection{Cross-model comparison: temporal evolution of AGN activity and star formation} \label{sec:cross_model}

\begin{figure*}
    \centering
    \includegraphics[width=\textwidth]{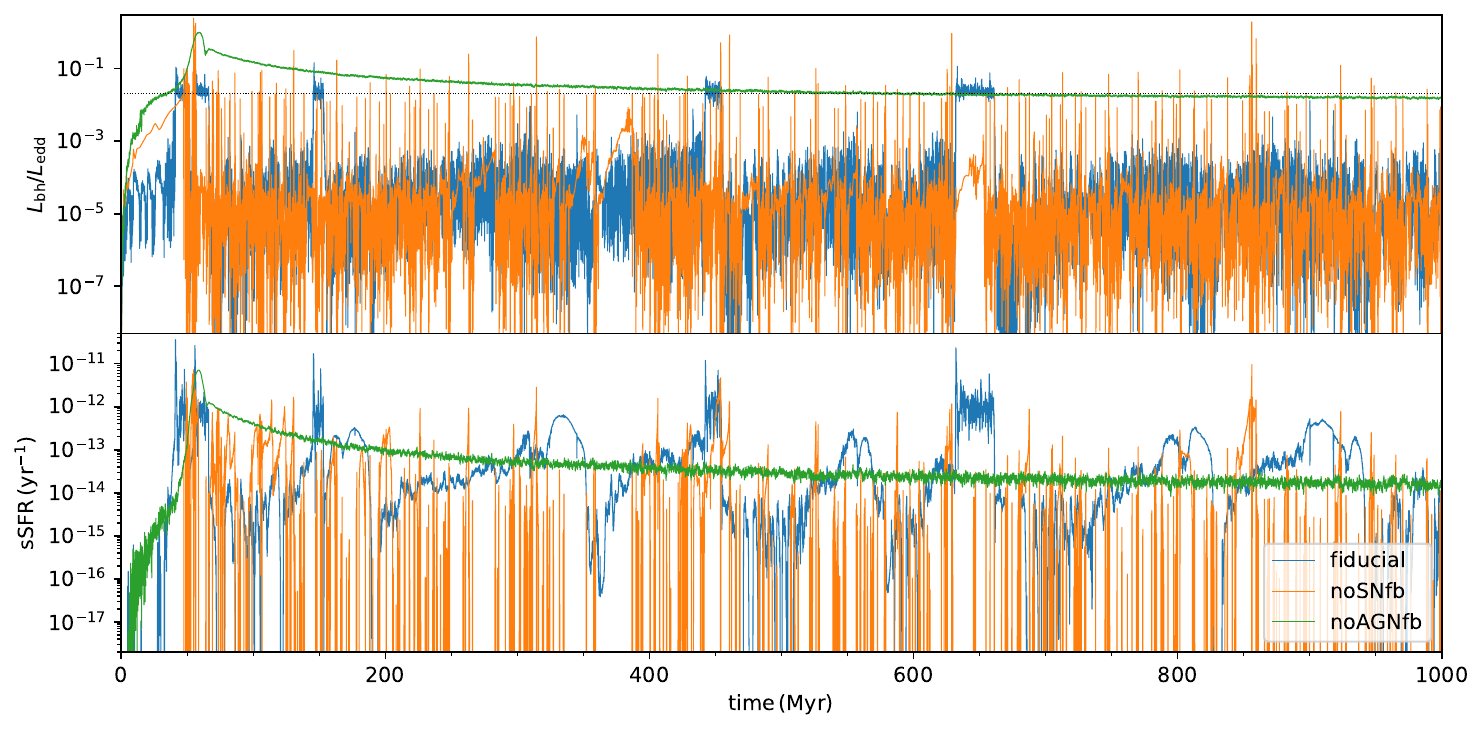}
    \caption{Temporal evolution of AGN luminosity $L_\mathrm{BH}$ (normalized by Eddington luminosity $L_\mathrm{Edd}$, top) and specific star formation rate sSFR (bottom) in the {\tt fiducial} (blue), {\tt noSNfb} (orange), and {\tt noAGNfb} (green) simulations. The dashed line in the top panel indicates $L_\mathrm{bh} / L_\mathrm{Edd} = 2\%$, delineating the transition between cold (above) and hot (below) AGN feedback modes. Each model exhibits distinctive evolutionary patterns: {\tt fiducial} shows correlated AGN-sSFR bursts at $\sim 10^2\,\mathrm{Myr}$ intervals, {\tt noSNfb} displays rapid fluctuations on $\sim$Myr timescales, and {\tt noAGNfb} maintains persistently elevated AGN activity.}
    \label{fig:lum_sfr_cmp_sims}
\end{figure*}

To investigate the relative impacts of AGN and SN feedback on galaxy evolution, we analyze the temporal evolution of AGN luminosity $L_\mathrm{BH}$ and specific star formation rate (sSFR) across our three simulation models (Fig.~\ref{fig:lum_sfr_cmp_sims}). Each model exhibits distinctly different evolutionary patterns, highlighting the complex interplay between these feedback mechanisms.

In the {\tt fiducial} simulation, which incorporates both feedback channels, AGN luminosity and sSFR demonstrate strong temporal correlation. The evolution is characterized by episodic bursts of activity persisting for $\sim \mathrm{Myr}$ with characteristic intervals of several hundred Myr. This behavior suggests coordinated regulation of both black hole accretion and star formation through the combined effects of AGN and SN feedback.

The {\tt noSNfb} simulation exhibits markedly different behavior, with both quantities showing rapid variability on $\sim$Myr timescales. The burst duration and intervals are substantially compressed compared to the {\tt fiducial} model. This pattern closely resembles the AGN luminosity evolution observed in previous MACER2D simulations where SN feedback was spatially smoothed, indicating that spatially-resolved SN feedback plays a crucial role in modulating AGN activity. The underlying mechanism likely involves SN feedback's influence on small-scale cold gas accretion dynamics, though detailed investigation of this process is deferred to future work.

The {\tt noAGNfb} simulation maintains consistently elevated AGN luminosity at $L_\mathrm{BH}/L_\mathrm{Edd} \sim 10^{-2}$ throughout the simulation period. While the sSFR remains predominantly below $10^{-12}\,\mathrm{yr^{-1}}$, suggesting apparent quiescence, this result requires careful interpretation. The simulation setup includes a central SMBH that functions as a gas sink through accretion but provides no energetic feedback. This artificial configuration may substantially underestimate the true star formation potential, as gas that would otherwise participate in star formation is continuously depleted by the SMBH. In a more realistic scenario lacking a central SMBH, gas accumulation in central regions could potentially drive significant star formation activity, potentially transitioning the galaxy into an actively star-forming state.

\subsection{Duty cycle of AGN activity} \label{sec:duty_cycle}

\begin{table}
    \centering
    \begin{tabular}{cccc}
        \hline
        Model & Duty cycle & Single-cycle timescale \\
        \hline
        {\tt fiducial} & 5.8\% & $275.2\,\mathrm{Myr}$ \\
        {\tt noSNfb} & 0.43\% & $2.7\,\mathrm{Myr}$ \\
        {\tt noAGNfb} & 100\% & $> 1\,\mathrm{Gyr}$ \\
        \hline
    \end{tabular}
    \caption{Comparison of the AGN duty cycles across simulation models, where the duty cycle is the fraction of time spent in active phases, and the averaged single-cycle timescale is the total duration of an active phase and the following inactive phase.
    Here, the criterion for an active phase is $L_\mathrm{bh} > 1\% L_\mathrm{Edd}$. The {\tt fiducial} model demonstrates characteristics of both duty cycle and single-cycle timescale consistent with observational constraints, while {\tt noSNfb} and {\tt noAGNfb} models exhibit significantly shorter and longer duty cycles and single-cycle timescale, respectively.}
    \label{tab:duty_cycle}
\end{table}

The distinct temporal evolution patterns of AGN luminosity and specific star formation rate across our three simulations manifest in remarkably different AGN duty cycles, as summarized in Table~\ref{tab:duty_cycle}. Here, the duty cycle represents the fraction of time the AGN spends in active phases (defined by $L_\mathrm{bh} > 1\% L_\mathrm{Edd}$), while the single-cycle timescale indicates the average duration of a complete cycle consisting of one active phase and its following inactive phase\footnote{In the {\tt fiducial} simulation's high-accretion stages, despite minor fluctuations where luminosity briefly drops below $1\% L_\mathrm{Edd}$, the AGN luminosity predominantly maintains $\sim 2\% L_\mathrm{Edd}$. Therefore, we consider each high-accretion stage as a single active phase.}. To minimize the influence of initial transients, our analysis considers only cycles occurring after $0.1\,\mathrm{Gyr}$ in each simulation.

The {\tt fiducial} simulation exhibits a duty cycle of 5.8\% with a single-cycle timescale of $275.2\,\mathrm{Myr}$. In contrast, the {\tt noSNfb} simulation demonstrates a much lower duty cycle of 0.43\% with substantially shorter cycles of $2.7\,\mathrm{Myr}$, indicating more rapid cycling with briefer active phases. The {\tt noAGNfb} simulation maintains persistently high AGN luminosity throughout the simulation period, resulting in a 100\% duty cycle and a single-cycle timescale exceeding the simulation duration. These results can be evaluated against observational constraints: multiple studies indicate typical AGN duty cycles of several percent with characteristic timescales of order $10^2\,\mathrm{Myr}$ \citep{greene2007mass,ho2009radiatively,kauffmann2009feast,conroy2012simple}. The {\tt fiducial} model's characteristics align well with these observational constraints, while the dramatically different duty cycle patterns in {\tt noSNfb} and {\tt noAGNfb} models suggest they do not accurately represent AGN activity in real galaxies.

\subsection{Mass growth of the black hole} \label{sec:mbh}

\begin{figure}
    \centering
    \includegraphics[width=\linewidth]{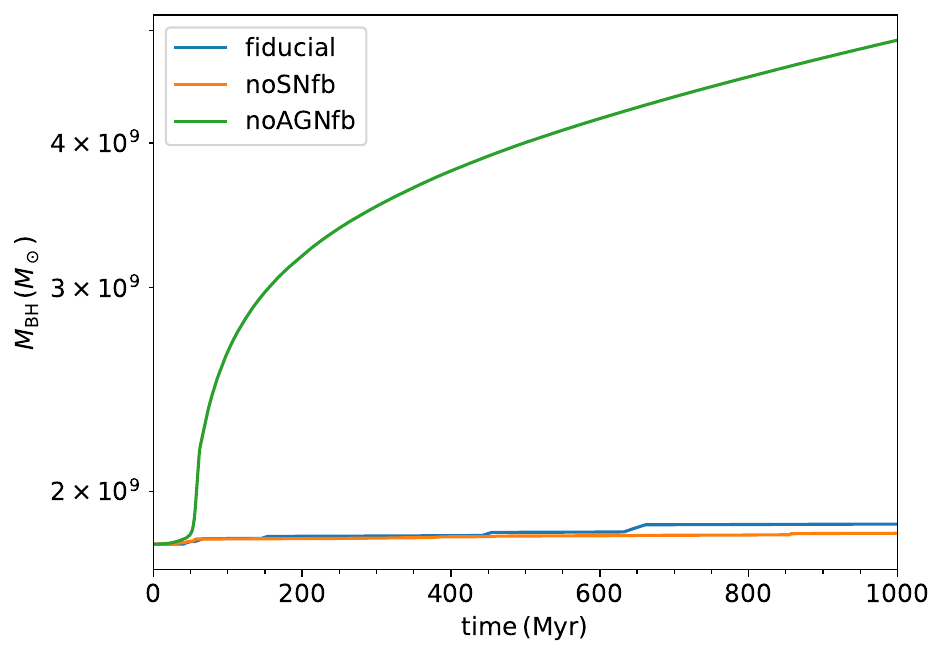}
    \caption{Time evolution of the black hole mass $M_\mathrm{BH}$ in the {\tt fiducial}, {\tt noSNfb}, and {\tt noAGNfb} simulations. The black hole mass grows steadily in all three cases, with the {\tt noAGNfb} model exhibiting the most rapid growth.}
    \label{fig:mbh}
\end{figure}

Fig.~\ref{fig:mbh} shows the time evolution of the black hole mass $M_\mathrm{BH}$ in the {\tt fiducial}, {\tt noSNfb}, and {\tt noAGNfb} simulations. The {\tt noAGNfb} simulation, due to the lack of AGN feedback, fails to expel gas from the central regions, leading to a continuous accretion of gas onto the BH. As a result, {\tt noAGNfb} exhibits the most rapid growth in $M_\mathrm{BH}$, with the BH mass increasing by $172\%$ over the course of the simulation. On the other hand, the {\tt fiducial} and {\tt noSNfb} simulations, which include AGN feedback, show more moderate growth in $M_\mathrm{BH}$, with the BH mass increasing by $5.7\%$ and $2.2\%$, respectively. The BH mass growth in different simulations is consistent with the AGN duty cycle and the AGN luminosity patterns, e.g., the BH growth in {\tt fiducial} shows a step-like pattern, with each step corresponding to a burst of AGN activity, while the BH growth in {\tt noAGNfb} is more continuous and rapid. The results suggest that AGN feedback plays a crucial role in regulating the BH growth, preventing the BH from growing too rapidly and maintaining the galaxy in a quiescent state.

A notable finding is that the BH mass growth in {\tt noSNfb} is slightly lower than in {\tt fiducial}, despite the absence of SN feedback in the former. This result suggests that SN feedback in the {\tt fiducial} simulation may enhance gas accretion onto the BH by driving turbulence and facilitating turbulent compression and mixing-induced cooling in the ISM, which promotes the formation of cool gas that fuels BH accretion. In contrast, the {\tt noSNfb} simulation with AGN feedback alone maintains an almost axisymmetric gas distribution with significantly reduced turbulence, resulting in less efficient cooling. Our analysis confirms that the gas cooling rate within the central region ($\lesssim 5\,\mathrm{pc}$) in {\tt noSNfb} is approximately two orders of magnitude lower than in {\tt fiducial}, leading to reduced cool gas formation and consequently lower BH accretion. This apparent ``positive SN feedback'' effect on BH growth through turbulence generation warrants further investigation in a dedicated study. However, it is important to note that this positive effect of SN feedback is observed when operating synergistically with AGN feedback, and it remains unclear whether SN feedback would maintain this positive influence on BH growth in the absence of AGN feedback, though the latter scenario seems unlikely.

\subsection{Star formation}

\begin{figure}
    \centering
    \includegraphics[width=0.8\linewidth]{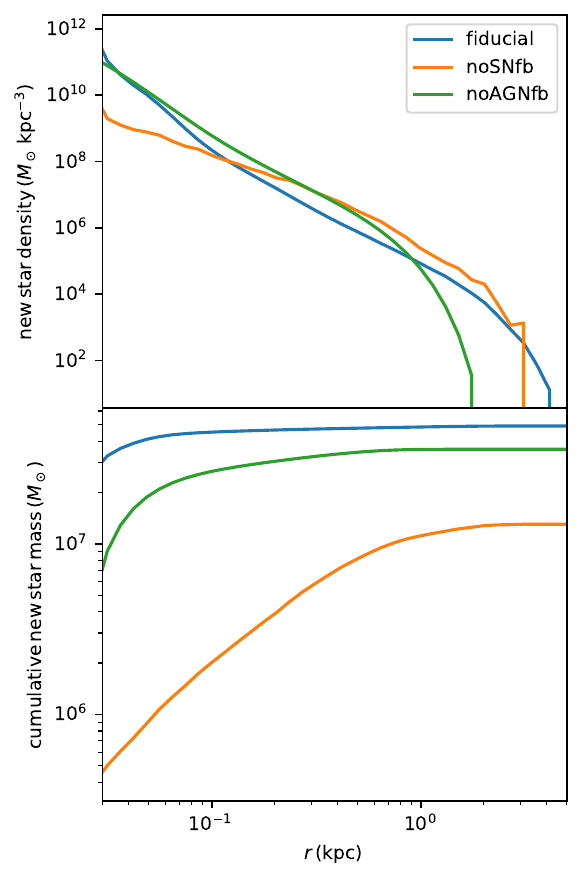}
    \caption{Profiles of the mass density of newly-formed stars (top) and the spatially cumulative mass of newly-formed stars from small to large radii (bottom) in the {\tt fiducial} (blue), {\tt noSNfb} (orange), and {\tt noAGNfb} (green) simulations at $t = 1000\,\mathrm{Myr}$. All three simulations exhibit centrally concentrated star formation, with total new stellar mass significantly below the galaxy's initial stellar content, indicating maintenance of quiescence. The {\tt fiducial} simulation demonstrates the highest star formation efficiency among the three models.}
    \label{fig:new_star}
\end{figure}

Fig.~\ref{fig:new_star} presents the radial distribution of newly-formed stellar mass density (top) and the cumulative mass of new stars as a function of radius (bottom) for our three simulation models. The stellar mass density profiles exhibit peak values at $r \sim 1\,\mathrm{kpc}$, with comparable magnitudes in the {\tt fiducial} and {\tt noAGNfb} simulations that exceed the {\tt noSNfb} peak by approximately two orders of magnitude. Both {\tt fiducial} and {\tt noAGNfb} models demonstrate steeper central density gradients compared to {\tt noSNfb}, indicating more spatially concentrated star formation. Notably, star formation activity is confined to regions within $r \sim 5\,\mathrm{kpc}$ across all simulations.

The cumulative mass distributions reveal steeper growth with radius in the {\tt fiducial} and {\tt noAGNfb} simulations relative to {\tt noSNfb}, consistent with their mass density profiles. The total mass of newly-formed stars reaches $4.9\times 10^7\,\mathrm{M_\odot}$, $3.6\times 10^7\,\mathrm{M_\odot}$, and $1.3\times 10^7\,\mathrm{M_\odot}$ in the {\tt fiducial}, {\tt noAGNfb}, and {\tt noSNfb} simulations respectively -- all negligible fractions of the galaxy's initial stellar mass. Remarkably, although still maintaining galaxy quiescence, the {\tt fiducial} simulation incorporating both feedback mechanisms exhibits the highest star formation efficiency. While this might suggest positive feedback effects, as discussed for SN feedback in \S\ref{sec:mbh}, it cannot be concluded that AGN feedback's impact on star formation is also positive: in the {\tt noAGNfb} simulation, the presence of an accreting central SMBH effectively deplete the gas reservoir potentially available for star formation (see \S\ref{sec:cross_model}), thus leading to smaller total new stellar mass compared to the {\tt fiducial} simulation. This does not imply the AGN feedback itself promotes star formation, but rather that the presence of an accreting BH without feedback can suppress star formation by depleting the gas reservoir more rapidly.

\subsection{Metal enrichment}

\begin{figure}
    \centering
    \includegraphics[width=\linewidth]{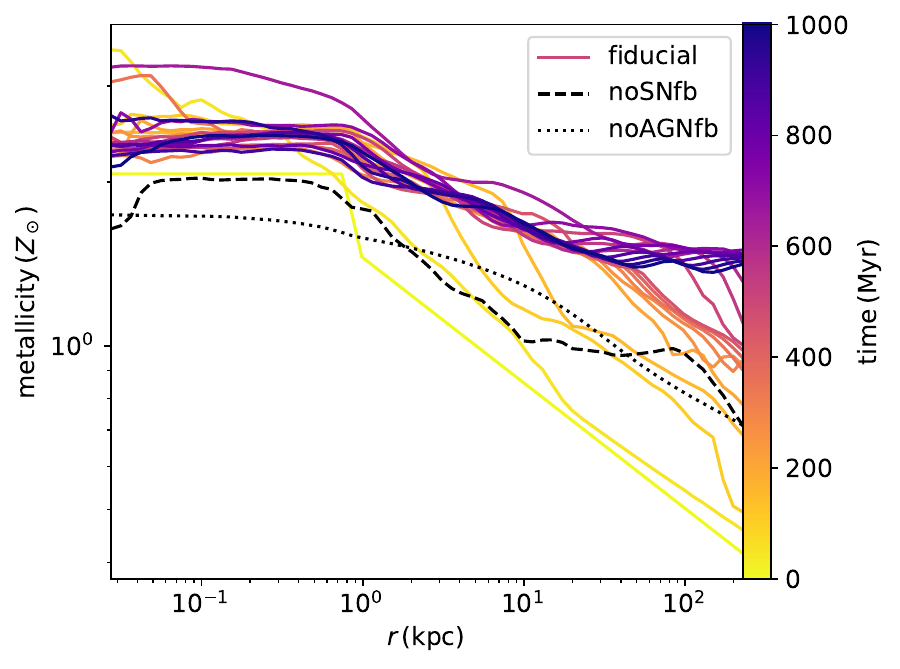}
    \caption{Radial metallicity profiles in the {\tt fiducial} simulation at different epochs from $t = 0\,\mathrm{Myr}$ to $1000\,\mathrm{Myr}$, with color gradient indicating simulation time. The metallicity profiles at $t = 1000\,\mathrm{Myr}$ from {\tt noSNfb} (dashed) and {\tt noAGNfb} (dotted) simulations are overlaid for comparison. In the {\tt fiducial} model, metallicity profiles evolve temporally and approach saturation near $t \sim 500\,\mathrm{Myr}$, with regions beyond $1\,\mathrm{kpc}$ exhibiting substantially greater enhancement compared to single-feedback models.}
    \label{fig:metallicity}
\end{figure}

Fig.~\ref{fig:metallicity} presents the color-coded temporal evolution of radial metallicity profiles in the {\tt fiducial} simulation from $t = 0\,\mathrm{Myr}$ to $1000\,\mathrm{Myr}$, alongside final metallicity profiles ($t = 1000\,\mathrm{Myr}$) from the {\tt noSNfb} (dashed) and {\tt noAGNfb} (dotted) simulations. In the {\tt fiducial} model, metallicity increases progressively with time until reaching approximate saturation around $t \sim 500\,\mathrm{Myr}$. While central regions ($r \lesssim 1\,\mathrm{kpc}$) demonstrate modest metallicity enhancement, outer regions ($r \gtrsim 1\,\mathrm{kpc}$) exhibit substantially greater enrichment, achieving slightly super-solar metallicity at CGM scales by the simulation's conclusion -- an enhancement of approximately one order of magnitude relative to initial conditions.

The {\tt noSNfb} and {\tt noAGNfb} simulations yield similar metallicity profiles that are systematically lower by factors of several compared to the {\tt fiducial} model beyond $1\,\mathrm{kpc}$. While observed CGM metallicities span a broad range up to $\sim 10\,[\mathrm{O/H}]$ \citep{zahedy2019characterizing}, accommodating predictions from all three models, our results suggest that the combined action of SN and AGN feedback substantially enhances CGM metal enrichment relative to either mechanism in isolation. This enhanced enrichment likely results from more efficient metal ejection from central regions followed by improved mixing and turbulent diffusion throughout the CGM when both feedback channels operate simultaneously. Future implementation will enable the evolution and tracking of individual metal species, providing a more detailed understanding of metal enrichment processes in galaxy evolution.

\subsection{X-ray properties of the gas} \label{sec:xray}

We analyze the soft X-ray emission from the {\tt fiducial} simulations by computing the radiation from collisional ionization equilibrium under the optically thin approximation, using the Astrophysical Plasma Emission Code (APEC) model \citep{smith2001collisional,foster2012updated}. Our calculations consider only gas emission, excluding contributions from the AGN or other point sources. The X-ray luminosity exhibits temporal variations spanning approximately two orders of magnitude, with peaks coinciding with AGN outbursts and a time-averaged value of $\sim 2.6\times 10^{41}\,\mathrm{erg\,s^{-1}}$. While this luminosity falls within observational constraints, it is several factors below the best-fit values from ROSAT and recent eROSITA observations \citep{anderson2015unifying,zhang2024circumgalactic}. We hypothesize that incorporating cosmological inflows, currently absent in our isolated galaxy simulations, might help increase the X-ray luminosity.

\begin{figure*}
    \centering
    \includegraphics[width=\textwidth]{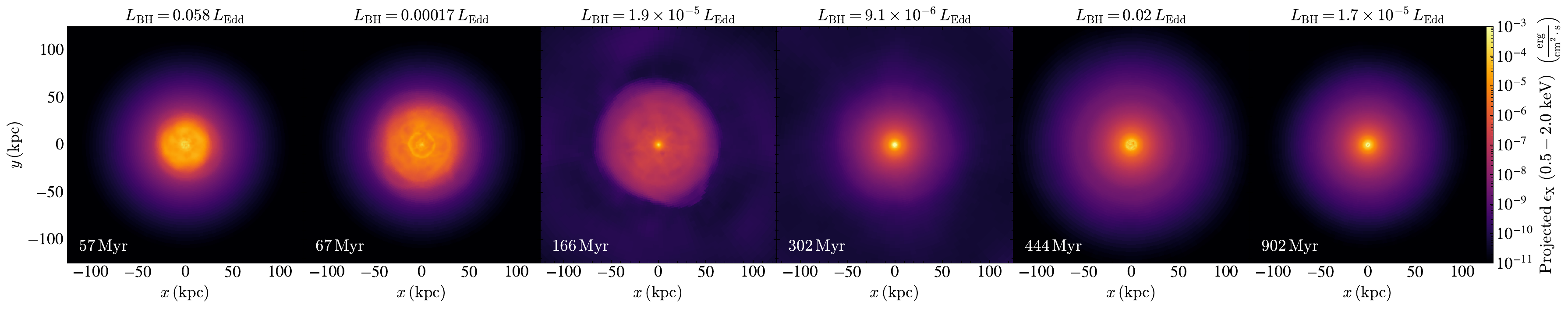}
    \includegraphics[width=\textwidth]{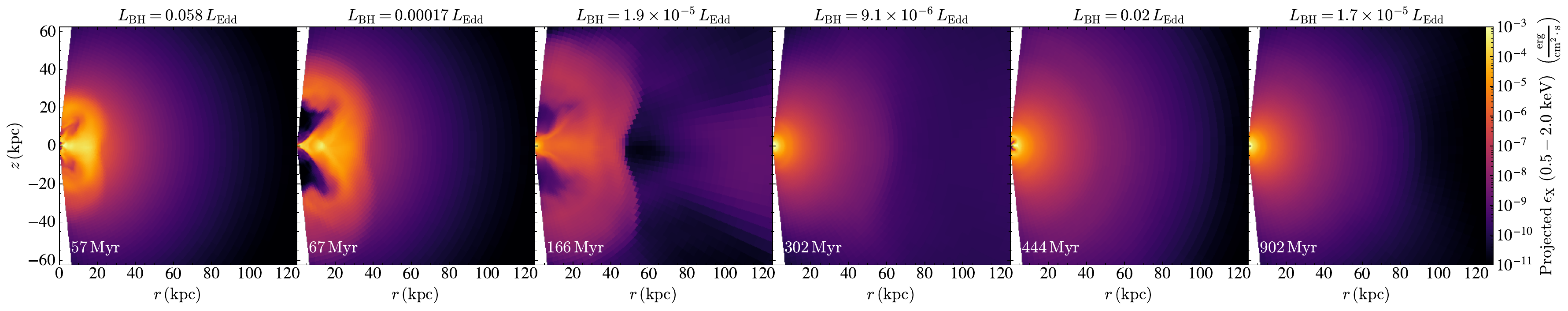}
    \caption{Projection plots of the X-ray luminosity viewed from the polar (top) and azimuthal (bottom) angles, zoomed in to a radius of $100\,\mathrm{kpc}$. The X-ray properties of the gas exhibit significant variations with time and viewing angles. X-ray cavities in size up to $20\,\mathrm{kpc}$ and bi-polar bubble structures are visible at certain times, which are caused by past cold-mode AGN feedback activities.}
    \label{fig:xray}
\end{figure*}

Fig.~\ref{fig:xray} shows the projection plots of the soft X-ray luminosity viewed from the polar (top) and azimuthal (bottom) angles, zoomed in to a radius of $100\,\mathrm{kpc}$, at different stages of the evolution as shown in Fig.~\ref{fig:theta_100} etc. The X-ray properties viewed from the polar angle are approximately spherical, while large-scale asphericity arises when viewed from the azimuthal angle at certain times, e.g., $t = 57\,\mathrm{Myr}$, $67\,\mathrm{Myr}$, and $166\,\mathrm{Myr}$. For instance, at $t = 67\,\mathrm{Myr}$, a $\sim 20\,\mathrm{kpc}$ X-ray cavity surrounded by a bi-polar bubble structure is visible, when the instantaneous AGN luminosity is low ($L_\mathrm{BH} \sim 1.7\times 10^{-4} L_\mathrm{Edd}$). The cavity is not produced by  AGN jets in the hot mode (AGN jets are not implemented yet in the simulation), but rather by high-speed ($\sim 10^4\,\mathrm{km}$) cold winds of the past AGN activity when the SMBH stays at high accretion rates, as shown in \S\ref{sec:gas_small_scale}. 

Two implications can be drawn from the results: (i) the large-scale X-ray properties, as the gas densities and temperatures, are not directly correlated with the instantaneous AGN luminosity but rather reflect the delayed AGN activities; (ii) X-ray cavities and bubble structures can be produced by past cold-mode AGN winds, albeit the size of the cavities is smaller than those created by AGN jets in the hot mode. We also note that the cavities are transient since the AGN active fraction is only a few percent (see \S\ref{sec:duty_cycle}). We suspect that it might be challenging to both detect these cavities produced by cold AGN winds due to their small sizes and short lifetimes, and to distinguish them from those produced by AGN jets in the hot mode. At low accretion rates ($t = 302\,\mathrm{Myr}$ and $902\,\mathrm{Myr}$), the X-ray properties are less aspherical. At $t = 414\,\mathrm{Myr}$, although the AGN luminosity is high at $\sim 0.02\,L_\mathrm{Edd}$, the spherical distribution of X-ray is also maintained, since the cold AGN winds just begin to launch (the cold wind is just visible at the very center in Fig.~\ref{fig:xray}) and have not yet influenced the large-scale X-ray properties. Future X-ray missions, such as XRISM \citep{2020XRISM} and HUBS \citep{2020HUBS,bregman2023scientific}, may provide valuable insights into the X-ray properties of the gas in galaxies and the impact of AGN feedback on the circumgalactic medium.

\section{Conclusions and Discussions} \label{sec:discussion}

\subsection{Conclusions}

We present the first results from the MACER3D framework, a new suite of three-dimensional hydrodynamic simulations of galaxy evolution featuring self-consistent two-mode AGN feedback and substantially enhanced gas physics and subgrid models compared to its MACER2D predecessor. Key improvements include a spatially-resolved SN feedback model with Poisson-distributed events, an exact integration scheme for gas heating and cooling, and a comprehensive metal enrichment model with spatially-resolved production, as detailed in \S\ref{sec:methods}. The simulations span a broad dynamical range from the fully-resolved vicinity of the Bondi radius ($25\,\mathrm{pc}$) to galactic halo scales ($\sim 250\,\mathrm{kpc}$), evolving over Gyr timescales. This extensive coverage proves crucial for accurately treating both small-scale accretion flows, which determine BH growth and AGN feedback modes, and capturing the long-term evolution of large-scale feedback dynamics simultaneously.

As the initial application of this framework, we simulate an isolated elliptical galaxy under three configurations: the {\tt fiducial} model with both SN and AGN feedback active, and two variants - {\tt noSNfb} and {\tt noAGNfb} - with either AGN or SN feedback disabled, respectively. These simulations exhibit distinct characteristics in AGN luminosity, star formation rate, AGN duty cycle, BH growth, star formation, metal enrichment, and X-ray properties. The distinct behaviors of these models provide valuable insights into the complex interplay between AGN and SN feedback mechanisms in regulating galaxy evolution, in particular, emphasizing the importance of SN feedback which is often underappreciated in elliptical environments.

The principal findings are:

\begin{enumerate}[label=(\roman*)]
    \item \emph{Positive temporal correlation between AGN luminosity and sSFR} The AGN luminosity and specific star formation rate in the {\tt fiducial} model demonstrate strong temporal correlation despite fluctuating across multiple orders of magnitude. Episodes of high-accretion cold mode AGN feedback ($L_\mathrm{bh} / L_\mathrm{Edd} > 2\%$) coincide with elevated sSFR ($\sim 10^{-12}\,\mathrm{yr^{-1}}$), while hot mode feedback maintains low sSFR ($\lesssim 10^{-13}\,\mathrm{yr^{-1}}$). This correlation indicates shared dependence on the available gas reservoir, while the galaxy maintains an overall quiescent state.

    \item \emph{Gas properties: instantaneous impact on small scales} Gas near the Bondi radius ($r \sim 1\,\mathrm{kpc}$) exhibits complex structures and dynamics that closely track instantaneous AGN luminosity. High-density cool gas inflows and high-velocity ($\sim 10^4\,\mathrm{kpc}$) cold winds manifest during high accretion periods, while clumpy or spiral-structured cool filaments with turbulent motions characterize low accretion states. At intermediate scales ($r \sim 5\,\mathrm{kpc}$), bi-conical structures triggered by past cold-mode AGN feedback are visible.

    \item \emph{Gas properties: delayed response on large scales} Gas properties at halo scales ($r \gtrsim 100\,\mathrm{kpc}$) reflect delayed AGN feedback effects rather than instantaneous luminosity. High-temperature shells with strong density and temperature gradients indicate past AGN-driven outflows, though no direct correlation exists between instantaneous AGN luminosity and halo-scale properties.

     \item \emph{Distinct duty cycles: SN feedback crucial for gas accretion regulation} The AGN duty cycles vary qualitatively across simulations. The {\tt fiducial} model exhibits observationally consistent duty cycles of several percent with a single-cycle timescale of $\sim 10^2\,\mathrm{Myr}$. Without SN feedback, the {\tt noSNfb} model shows sub-percent duty cycles and a single-cycle timescale of $\sim$Myr, highlighting SN feedback's crucial role in regulating gas accretion and thus AGN activity. The {\tt noAGNfb} model maintains consistently high AGN luminosity without clear cycling.

    \item \emph{Feedback effects on BH growth and star formation: negative but complex} AGN feedback effectively constrains BH growth, limiting mass increase to a few percent compared to nearly 200\% growth without feedback. Star formation remains suppressed across all models, with newly-formed stellar mass reaching only $\sim 10^7\,\mathrm{M_\odot}$. Notably, the {\tt fiducial} model exhibits several-fold higher star formation efficiency than single-feedback models, suggesting potentially positive SN feedback effects through turbulence-enhanced gas mixing and cooling. Lower star formation in {\tt noAGNfb} is due to the central SMBH accreting gas at a high rate, reducing the gas available for star formation, rather than the AGN feedback itself is positive for star formation.

    \item \emph{Enhanced metal enrichment through combined feedback} The {\tt fiducial} model achieves super-solar CGM metallicity by $t \sim 500\,\mathrm{Myr}$, exceeding single-feedback models by factors of several beyond $1\,\mathrm{kpc}$, while maintaining comparable central metallicity. This suggests synergistic enhancement of CGM metal enrichment through combined feedback mechanisms.

    \item \emph{X-ray properties: delayed response with transient asphericity} Hot gas X-ray luminosity varies by two orders of magnitude, averaging $\sim 10^{41}\,\mathrm{erg\,s^{-1}}$ with peaks coinciding with AGN outbursts. While primarily spherical, X-ray emission shows transient $\sim 20\,\mathrm{kpc}$ cavities and bubble structures triggered by previous cold-mode AGN winds, demonstrating delayed response to AGN activity.
\end{enumerate}

\subsection{Comparison with results from MACER2D}

The MACER3D simulations presented here build upon the MACER2D framework, which has been validated and applied to study AGN feedback in galaxy evolution. Some key results obtained under the MACER3D framework are consistent with those from MACER2D simulations, including the suppression of star formation and black hole growth by AGN feedback, the heavily time-variable AGN luminosity, etc. Nevertheless, a few qualitative differences in the results between the MACER3D and MACER2D simulations are worth noting.

First, the increase of dimensionality allows for a more comprehensive and realistic treatment of gas dynamics, fluid and thermal instabilities, and turbulence. In MACER2D simulations, the gas dynamics are inherently axisymmetric, while MACER3D simulations capture complex three-dimensional structures and dynamics which are particularly important for accurately modeling gas accretion. For instance, non-axisymmetric structures such as cool filaments spiraling into the central regions are observed in MACER3D simulations, but cannot exist in 2D because of the axisymmetry. In addition, in MACER2D simulations, long-standing large-scale eddies are constantly presented, while small-scale gas structures are absent given even higher spatial resolution in 2D simulations. This is due in great part to the inverse cascade of turbulence from small to large scales in two dimensions. In contrast, in the MACER3D simulations where turbulence cascades from large to small scales, such artificial, long-lasting large-scale eddies are not observed, and small-scale structures are more prominent.

Second, the MACER3D simulations exhibit significantly different AGN duty cycles compared to MACER2D simulations. Each AGN duty cycle in MACER3D simulations lasts for $\sim 10^2\,\mathrm{Myr}$, unambiguously consisting of one active phase of $\sim 10\,\mathrm{Myr}$ and one quiescent phase for the rest of the cycle. The single-cycle timescale in MACER2D simulations is much shorter, with the AGN luminosity fluctuating rapidly between active and quiescent phases on $\sim$Myr or shorter timescales, without clear long-lasting active or quiescent phases. Since the {\tt noSN} simulations in MACER3D exhibit similar rapid fluctuations in AGN luminosity as in MACER2D simulations, we suspect that the spatially-resolved SN feedback in MACER3D simulations plays a crucial role in modulating AGN activity, which is absent in MACER2D simulations.

Moreover, since the single-cycle timescale is much longer in MACER3D simulations, the temporal correlation between AGN luminosity and sSFR is pronounced in MACER3D simulations (refer to Fig.~\ref{fig:lum_sfr}), suggesting a shared dependence on the available gas reservoir. This relationship, although it might exist in MACER2D simulations, is not so apparent due to the much shorter single-cycle timescale. The detailed investigation of the related underlying physical mechanisms, particularly whether turbulence and radiative cooling cooperating with feedback channels modulate the gas supply for both black hole accretion and star formation, will be the focus of a subsequent study.

Finally, the MACER3D framework is designed to be more general and flexible, and can be applied to a wide range of galaxy evolution studies, including disk galaxies, dwarf galaxies, etc., and can be extended to include more physical processes, such as magnetic fields and cosmic rays (see \S\ref{sec:future}). The axisymmetric nature of MACER2D simulations limits their comprehensive applicability to certain types of galaxies, such as disk galaxies, where three-dimensional effects, e.g., gravitational torques and non-axisymmetric instabilities, are indispensable for the angular momentum transport of the gas in the galaxy. 

Although the discussion above is dedicated to the comparison between MACER3D and MACER2D simulations, it reflects more general differences between 3D and 2D models, and between single-channel and multi-channel feedback implementations. The comparison suggests that three-dimensional modeling with comprehensive feedback physics is essential for accurately capturing galaxy evolution, even in relatively simple systems like isolated elliptical galaxies. While AGN feedback has long been recognized as crucial for maintaining quiescence in massive ellipticals, the comparison demonstrates that accurately modeling SN feedback is equally important, even in ellipticals.

\subsection{Caveats}

We acknowledge several important caveats and limitations of the current study:

\begin{enumerate}[label=(\roman*)]
    \item \emph{Limited CGM resolution} While our simulations achieve high resolution near the Bondi radius, which proves essential for accurately modeling gas accretion and feedback dynamics, the logarithmically decreasing resolution toward outer regions may inadequately capture thermal instabilities and turbulence in the CGM \citep{mccourt2017characteristic,peeples2018figuring,hummels2018impact}. The observed cool ($\sim 10^4\,\mathrm{K}$) gas component in the CGM of elliptical galaxies (e.g., \citealt{zahedy2019characterizing}) is significantly underrepresented in our simulations, potentially due to either unresolved thermal instabilities or the absence of cool cosmological inflows \citep{afruni2019cool}.

    \item \emph{Absence of cosmological context} Our focus on isolated elliptical galaxies excludes cosmological inflows, while inflows could enhance both X-ray luminosity (as noted in \S\ref{sec:xray}) and cool gas content in the CGM. While future work will incorporate these inflows, we anticipate that their impact on feedback physics may be limited, as \citet{zhu23a} demonstrated minimal penetration of inflows into galactic central regions, though this conclusion warrants verification in three dimensions.

    \item \emph{Exclusion of hot-mode AGN jets} The current implementation does not include AGN jets during hot-mode accretion, despite their established importance as feedback channels and their role in creating observed X-ray cavities \citep{2007McNamara,2014Heckman}. The effects of AGN jets within the MACER framework are currently under investigation (Guo et al., in prep).

    \item \emph{Subgrid model limitations} Our simulations necessarily rely on several subgrid prescriptions for processes including star formation and metal enrichment. While such approximations are inherent to galaxy evolution simulations, we have deliberately chosen simple implementations with minimal free parameters to capture essential physics while maintaining tractability. A comprehensive exploration of alternative subgrid models lies beyond our current scope but merits future investigation.

    \item \emph{Non-thermal physics} Our simulations do not yet include important non-thermal processes such as magnetic fields and cosmic rays, which may significantly influence feedback physics and gas dynamics, particularly in the low-pressure CGM environment (e.g., \citealt{butsky2018role,hopkins2020but,ji2020properties,buck2020effects,van2021effect}). AGN feedback likely drives small-scale dynamo action and magnetic field amplification, while AGN jets can accelerate cosmic rays. These effects will be incorporated in forthcoming work, including Xia et al. (in prep).
\end{enumerate}

\subsection{On going and future work} \label{sec:future}

In line with the main results presented above, a subsequent study will focus on the underlying physical mechanisms driving the observed properties in this study, particularly how turbulence and radiative cooling cooperating with feedback channels modulate the gas supply for both black hole accretion and star formation. Near-term developments of the MACER3D framework include implementing AGN jets in hot mode feedback (Guo et al., in prep) and incorporating magnetic fields (Xia et al., in prep). The framework will be extended to simulate disk galaxies (Zou et al., in prep) and dwarf galaxies (Su et al., in prep) to investigate AGN feedback across different galactic environments. Longer-term goals include implementing cosmic rays and other potentially important physics, improving CGM resolution, and incorporating cosmological inflows. These enhancements will enable more comprehensive studies of AGN feedback's role in galaxy evolution while maintaining the framework's idealized nature for controlled physical investigation.

\section*{Acknowledgments}

We dedicate this work to the memory of Professor Jeremiah P. Ostriker. The MACER model was developed step by step upon the foundation laid by his pioneering work and benefited greatly from his long-standing support and encouragement. In particular, FY is deeply grateful for his mentorship over the decades, which shaped much of his research path and especially inspired FY to enter the field of AGN feedback.

We thank the anonymous referee for their constructive comments and suggestions that helped improve the quality of this paper. We also thank Fangzheng Shi for very helpful discussions. Authors are supported by the Natural Science Foundation of China (grants 12133008, 12192220, 12192223, and 12361161601), the China Manned Space Program through its Space Application System, and the National Key R\&D Program of China No. 2023YFB3002502. This work was performed in part at the Aspen Center for Physics, which is supported by National Science Foundation grant PHY-2210452. The data supporting the plots within this article are available on reasonable request to the corresponding author. Numerical calculations were run on the CFFF platform of Fudan University, the supercomputing system in the Supercomputing Center of Wuhan University, and the High Performance Computing Resource in the Core Facility for Advanced Research Computing at Shanghai Astronomical Observatory. We have made use of NASA's Astrophysics Data System.

\vspace{5mm}

\software{{\small Athena++} \citep{Athenapp2020},
          {\small Cloudy} \citep{cloudy},
          {\small Matplotlib} \citep{hunter2007matplotlib},
          {\small NumPy} \citep{2020NumPy-Array}, 
          {\small SciPy} \citep{2020SciPy-NMeth}, 
          {\small yt} \citep{Turk2010,turk2024introducing}
          }


\bibliography{ref}
\bibliographystyle{aasjournal}
\end{document}